\documentclass[aps,10pt,prb,twocolumn,groupedaddress,showpacs]{revtex4}

\usepackage{graphicx}
\usepackage[caption=false]{subfig}
\usepackage{amsmath}
\usepackage{bbold}
\usepackage{latexsym}

\begin{document}


\title{Influence of device geometry on tunneling in $\nu=\frac{5}{2}$ quantum Hall liquid}


\author{Guang Yang and D. E. Feldman}
\affiliation{Department of Physics, Brown University, Providence, Rhode Island 02912, USA}


\date{\today}

\begin{abstract}
Two recent experiments [I. P. Radu {\it et al.}, Science $\mathbf{320}$, 899 (2008) and
X. Lin {\it et al.}, Phys. Rev. B $\mathbf{85}$, 165321 (2012)] measured the temperature and voltage dependence of the quasiparticle tunneling through a quantum point contact in the $\nu= 5/2$ quantum Hall liquid. The results led to conflicting conclusions about the nature of the quantum Hall state. In this paper, we show that the conflict can be resolved by recognizing different geometries of the devices in the experiments. We argue that in some of those geometries there is significant unscreened electrostatic interaction between the segments of the quantum Hall edge on the opposite sides of the point contact. 
Coulomb interaction affects the tunneling current. We compare experimental results with theoretical predictions for the Pfaffian, $SU(2)_2$, $331$ and $K=8$ states and their particle-hole conjugates. After Coulomb corrections are taken into account, measurements in all geometries agree with the spin-polarized and spin-unpolarized Halperin 331 states.
\end{abstract}



\maketitle



\section{INTRODUCTION \label{sec:intro}}

\begin{figure}
\centering
\includegraphics[width=3in]{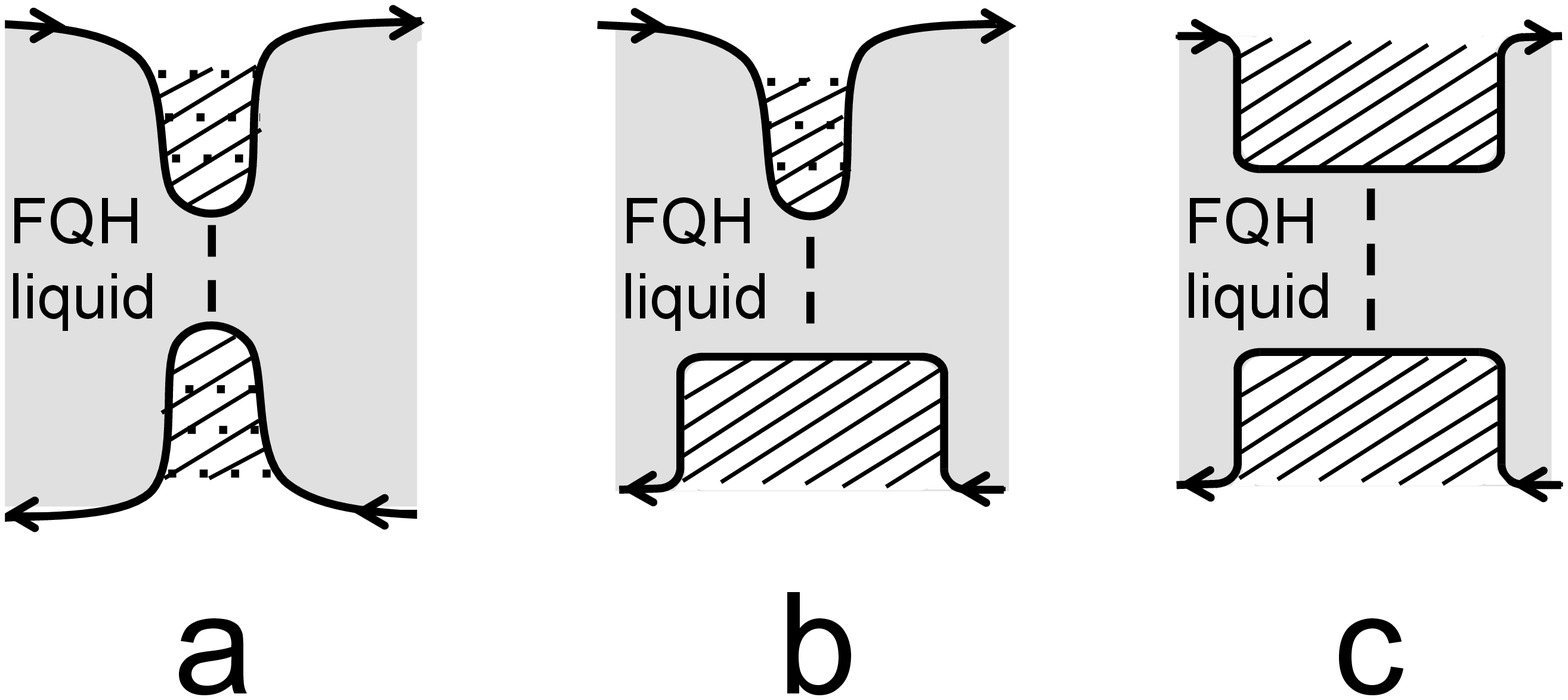}
\caption{The three QPC geometries in the experiments. The arrows follow the current propagation direction  on the edge.  Dashed lines denote quasiparticle tunneling. Dotted lines across narrow gates in the geometries 1a and 1b represent the electrostatic interaction.}
\end{figure}

Among numerous phases of two-dimensional electron gases (2DEG), the even-denominator quantum Hall states with the filling factors \cite{willett87} $5/2$ and $7/2$ are particularly interesting. In contrast to odd-denominator fractional quantum Hall (FQH) liquids, they cannot be explained by a straightforward generalization of the Laughlin variational wave function. An early attempt to understand their nature led to the beautiful idea of non-Abelian states of matter \cite{2}. In non-Abelian systems, the types and positions of quasiparticles do not uniquely determine the quantum state. This results in unusual physics and may open a road to topological quantum computing \cite{6,7}. 
However, the existence of non-Abelian quasiparticles has not been proven and the nature of the 5/2 state remains a puzzle. 

Both Abelian and non-Abelian candidate states were proposed as possible theoretical explanations of the 5/2 FQH effect \cite{2,3,4,5,8,19,BS,26,Overbosch}.  A number of methods 
\cite{Stern10,fradkin98,dassarma05,11,12,14,hou06,grosfeld06,rosenow12,feldman08,viola12,overbosch09,seidel09,wang10a,CS,yang09}
were invented and several experiments 
\cite{9,willett09,willett10,36,mstern10,rhone11,tiemann12,mstern12,chickering13,kang,20,21}
were performed in an attempt to determine the right ground state. One approach \cite{20,21} consists in the measurement of the tunneling current through a quantum point contact (QPC) between the edges of a 5/2 FQH liquid. The low-temperature conductance exhibits a power-law behavior $G\sim T^{2g-2}$, where the exponent $g$ depends on the topological order in the bulk \cite{wen}. The exponent $g$ was measured in two recent experiments \cite{20,21}. The results of the earlier experiment \cite{20} were interpreted as supporting the non-Abelian anti-Pfaffian or $SU(2)_2$ states. The best fit for the second experiment \cite {21} comes from the Abelian 331 state.

We argue that the discrepancy between those results can be explained by different geometries of point contacts in Refs. \onlinecite{20} and \onlinecite{21}. The geometry of the device from Ref. \onlinecite{20} is shown schematically in Fig. 1a. Figs. 1b and 1c illustrate the two geometries from Ref. \onlinecite{21}. In all cases, transport is supported by FQH edge channels defined by top gates. In Fig. 1a and for the upper gate in Fig. 1b, the gate widths are comparable to their distances from 2DEG. The gates are considerably wider in Fig. 1c. Thus, the distance between the edge points, connected by the dotted lines in Figs. 1a and 1b, is shorter than the screening length due to the metallic gates. In other words, the repulsive Coulomb interaction between the FQH edge segments on the opposite sides of QPC is not fully screened by the gates in the geometries Fig. 1a,b but is screened completely in the geometry of Fig. 1c. 

It is well known that repulsive interaction suppresses tunneling in quantum wires \cite{giamarchi}. Similar physics has been addressed in the context of edge transport in Laughlin FQH states in Refs.
\onlinecite{41},\onlinecite{papa05}. We find that in our case the Coulomb interaction drives the tunneling exponents $g_A$ and $g_B$ in the geometries 1a and 1b above the exponent $g_C$ in the geometry 1c. Only the latter assumes a universal value while the former depend on the strength of the interaction across the gates. Hence, information about the nature of the 5/2 state can be extracted from a comparison of the experimental $g_C$ (Ref. \onlinecite{21}) but not $g_A$ and $g_B$ with universal theoretical predictions in various proposed phases. Such predictions are available for the Abelian $K=8$ and spin-polarized and unpolarized $331$ states\cite{8,19,Overbosch}, the non-Abelian $SU(2)_2$ and Pfaffian states \cite{2,7,Overbosch} and the particle-hole conjugate of the Pfaffian state, the anti-Pfaffian state \cite{3,4}. Since Ref. \onlinecite{36} provides support for the existence of contra-propagating edge channels in the 5/2 FQH liquid, it is particularly interesting to consider candidate states with upstream edge modes. In the above list, only the anti-Pfaffian state has such property. The particle-hole conjugates of the other proposed states also exhibit upstream edge transport and we compute $g_C$ in them below. The comparison with the experiment shows that the Halperin 331 state agrees best with the tunneling data \cite{21}. 

This paper is organized as follows. We review theoretically proposed $\nu=5/2$ FQH states in Section~II. Three appendices complement that section. In Appendix A we discuss the edge theory of the anti-331 state which can be understood as the Bonderson-Slingerland state \cite{BS} built on the particle-hole conjugate of the Laughlin 1/3 state. Appendix B addresses the anti-$SU(2)_2$ state. In Appendix C, we argue that the anti-$K=8$ state does not exhibit a universal tunneling exponent. Effects of the Coulomb interaction across the gate are discussed qualitatively in Section III. Detailed calculations are given in Section IV and Appendix D. In the final section we compare our results and
the experimental findings \cite{20,21} with the information obtained from other types of experiments.

\section{Candidate $\nu=5/2$ FQH states }

Here we review several simplest topological orders, proposed for the $5/2$ state, and address their edge properties which we will need in the subsequent sections. 

In all candidate states, the lowest quasiparticle charge is $e/4$, where $e$ is the electron charge. This is rather different from the odd denominator states, where we usually expect that the lowest quasiparticle charge of the $p/q$
state is $e/q$. The $e/4$ charge follows a general rule \cite{levin09} for even denominator states: The ratio of  the quantized Hall conductance to the lowest possible charge must be $2ne/h$, where $n$ is an integer.
The $e/4$ value of the charge agrees with the experiments \cite{9,willett09,21} on the 5/2 FQH liquids.
The best fits for the quasiparticle charge in the tunneling experiments \cite{20,21} are $0.17e$ in the geometry 1a, $0.25e$ in the geometry 1b and $0.22e$ in the geometry 1c. 
In all cases the nearest possible excitation charge is $e/4$.
Thus, we conclude that in all geometries the tunneling current is carried by $e/4$-quasiparticles.

We start our overview with the simplest Abelian topological orders. To simplify our notation we set $\hbar$ to 1.

\subsection{$K=8$ state}

We assume that electrons form pairs in the second Landau level and single-electron excitations are gapped. In particular, no single-electron tunneling into
the FQH edge channel of  a $5/2$ system is possible at low energies. The filling factor for bosonic electron pairs is $1/8$. Thus, we use the edge theory of a Laughlin state of bosons at that filling factor. The Lagrangian density \cite{Overbosch}

\begin{equation}
\label{dima1}
\mathcal{L}=-\frac{1}{4\pi}[8\partial_t\phi\partial_x\phi+8v(\partial_x\phi)^2]+\mathcal{L}_{2}+\mathcal{L}_{\rm int},
\end{equation}
where 

\begin{equation}
\label{dima2}
\mathcal{L}_{2}=-\frac{1}{4\pi}[\sum_{i=1,2} \partial_t\phi^0_i\partial_x\phi^0_i+\sum_{ij}U_{ij}\partial_x\phi^0_i\partial_x\phi^0_j]
\end{equation}
describes two integer edge modes, $\mathcal{L}_{\rm int}$ contains information about the interaction of the integer and fractional modes,
the charge density on the FQH edge is $\rho(x)={e\partial_x\phi}/{\pi}$ and the charge density on the integer edges is $\rho_{2}={e(\phi^0_1+\phi^0_2)}/{2\pi}$. The operator $\Psi(x)=\exp(i\phi(x))$ annihilates a charge-$e/4$ anyon. The pair annihilation operator on the fractional edge is $\Psi_{2e}(x)=\exp(i 8\phi(x))$. 

We do not make assumptions about the spin of the electron pairs in the $K=8$ state. The same edge theory describes spin-polarized and unpolarized FQH liquids. Certainly, the particle-hole conjugate system has the filling factor 5/2 only for the spin-polarized $K=8$ liquid.

\subsection{331 state}

This state comes in two shapes with identical signatures in the tunneling experiment but different spin polarizations \cite{Overbosch}.

The spin-unpolarized version can be understood as a Halperin bilayer state with spin-up and -down electrons playing the role of two layers. 
The Lagrangian density on the edge

\begin{equation}
\label{dima3}
\mathcal{L}= -\frac{1}{4\pi}  \sum_{I,J=1,2} [K_{IJ}\partial_{t}\phi_{I}\partial_{x}\phi_{J}+V_{IJ}\partial_{x}\phi_{I}\partial_{x}\phi_{J}]+\mathcal{L}_{2}+\mathcal{L}_{\rm int},
\end{equation}
where the $K$-matrix\cite{wen}

\begin{equation}
\label{dima4}
K =K_{\rm unpolarized}=
\left( \begin{array}{cc}
3 & 1 \\
1 & 3 \end{array} \right)
\end{equation}
and $\mathcal{L}_{2}$ and $\mathcal{L}_{\rm int}$ are defined in the same way as in the previous subsection.
The charge density on the fractional edge is $e(\partial_x\phi_1+\partial_x\phi_2)/2\pi$ and the most relevant operators that annihilate quasiparticles with charge $e/4$ are $\exp(i\phi_{1,2})$. There are two independent electron operators: $\Psi_{1}(x)=\exp(i[3\phi_1+\phi_2])$ and $\Psi_2(x)=\exp(i[3\phi_2+\phi_1])$.
The spin-unpolarized 331 state is not an eigenstate of the total spin and is related \cite{23} to the physics of superfluid He-3.

The spin-polarized version of the 331 state emerges from the condensation of the charge-$2e/3$ quasiparticles on top of the Laughlin state with the filling factor 1/3. The $K$-matrix

\begin{equation}
\label{dima5}
K =K_{\rm polarized}=
\left( \begin{array}{cc}
3 & -2 \\
-2 & 4 \end{array} \right)
\end{equation}
can be expressed as $K_{\rm polarized}=W^{\rm T}K_{\rm unpolarized}W$ with 

\begin{equation}
\label{dima6}
W =
\left( \begin{array}{cc}
0 & 1 \\
1 & -1 \end{array} \right)
\end{equation}
and hence describes the same topological order as the matrix (\ref{dima4}). Note that the particle-hole conjugate of the 331 state has the filling factor 5/2 only for its spin-polarized version.

\subsection{Pfaffian state}

The non-Abelian Pfaffian state \cite{2,7} can be seen as a $p$-wave superconductor with the wave function 

\begin{equation}
\label{dima7}
{\rm Pf}[\frac{1}{z_i-z_j}]\Pi_{i<j}(z_i-z_j)^2\exp(-\frac{1}{4}\sum|z_i|^2),
\end{equation}
where $z_k=x_k+iy_k$ is the position of the $k$th electron, ${\rm Pf}$ stays for the Pfaffian of a matrix and we ignore the two filled Landau levels for a moment.
The edge theory is described by the following Lagrangian density

\begin{equation}
\label{dima8}
\mathcal{L}=-\frac{2}{4\pi}[\partial_t\phi_1\partial_x\phi_1+v_1(\partial_x\phi_1)^2] +i\psi(\partial_t+v_\psi\partial_x)\psi+\mathcal{L}_{2}+\mathcal{L}_{\rm int},
\end{equation}
where the charge density on the fractional edge is $e\partial_x\phi_1/2\pi$, $\psi$ is a neutral Majorana fermion, $v_1$ and $v_\psi$ are the velocities of the charged and neutral modes and the meaning of $\mathcal{L}_2$ and
$\mathcal{L}_{\rm int}$ is the same as in the preceding subsections. The charge $e/4$-quasiparticle annihilation operator $\Psi(x)=\sigma(x)\exp(i\phi_1(x)/2)$, where the operator $\sigma$ changes the boundary conditions for the Majorana fermion from periodic to antiperiodic and has the scaling dimension $1/16$. The Pfaffian state is spin-polarized.

\subsection{$SU(2)_2$ state}

This is another spin-polarized non-Abelian state. Its wave function can be derived from the parton construction \cite{5}. We split an electron into a fermion $\psi_{1/2}$ of charge $e/2$ and two fermions $\psi_{1/4a,}$, $\psi_{1/4,b}$
of charge $e/4$: 

\begin{equation}
\label{dima9}
\Psi_e=\psi_{1/2}\psi_{1/4,a}\psi_{1/4,b}.
\end{equation}
The filling factor for the $e/2$-partons is 1. For each of the two sorts of $e/4$-partons the filling factor is 2. For decoupled partons the wave function would simply be a product of three integer quantum Hall
(IQH) wave functions. Taking into account that the three types of partons occupy exactly the same positions, we find the ground state wave function of the form \cite{Overbosch}

\begin{equation}
\label{dima10}
\Psi(\{z_k\})=[\chi_2({z_k})]^2\Pi_{k<l}(z_k-z_l)\exp(-\frac{1}{4}\sum|z_k|^2),
\end{equation}
where $\chi_2$ is the polynomial factor in the wave function of two filled Landau levels.

The decomposition (\ref{dima9}) exhibits $U(1)\times SU(2)$ gauge symmetry (the factor $SU(2)$ acts on $\psi_{1/4,a}$ and $\psi_{1/4,b}$; $U(1)$ describes the freedom to choose the phase of $\psi_{1/2}$). For decoupled partons, the edge theory would contain 5 IQH edge channels. The theory of the FQH edge is obtained by keeping only gauge invariant states and is described by the Lagrangian density \cite{Overbosch}

\begin{align}
\label{dima11}
\mathcal{L}=  -\frac{1}{4\pi} & [2\partial_{t}\phi_{\rho}\partial_{x}\phi_{\rho}+\partial_{t}\phi_{n}\partial_{x}\phi_{n}+2v_{\rho} (\partial_{x}\phi_{\rho})^{2}  \nonumber\\
&+v_{n} (\partial_{x}\phi_{n})^{2}]+i \psi (\partial_{t}+v_{\psi}\partial_{x})\psi ,
\end{align}
where $\phi_{\rho}$ is a bosonic mode carrying electric charge, $\phi_{n}$ is an electrically neutral bosonic mode and $\psi$ is a neutral Majorana fermion. The charge density is $e\partial_x\phi_\rho/2\pi$. The neutral mode $\phi_n$
describes ``pseudo-polarization'', i.e., the difference in the occupation of the two Landau levels for $e/4$-partons.
The two integer edge modes should be included in the same way as in Eqs. (\ref{dima1},\ref{dima3},\ref{dima8}).
 The $e/4$-quasiparticle annihilation operators are $\sigma(x)\exp(\pm i\phi_n(x)/2)\exp(i\phi_\rho(x)/2)$, where $\sigma$ twists the boundary conditions for the Majorana fermion and has the scaling dimension 1/16. Three operators annihilate an electron on the fractional edge: $\psi(x)\exp(2i\phi_\rho(x))$ and $\exp(\pm i\phi_n(x))\exp(2i\phi_\rho(x))$. 

In the presence of disorder the ``pseudo-polarization'' does not conserve and the action can contain tunneling operators of the form $\xi_{\pm}(x)\psi(x)\exp(\pm i\phi_n(x))$, where $\xi_{\pm}(x)$
are random functions of the coordinate. Their effect is the same as the effect of similar operators in the anti-Pfaffian state \cite{3,4} and the anti-331 and anti-$SU(2)_2$ states (Appendices A and B). 
The theory acquires an emergent $SO(3)$ symmetry. The neutral modes should be described as three Majorana fermions with equal velocities. In contrast to Appendices A and B, however, this does not affect the scaling dimension of the most relevant quasiparticle operators.

\subsection{Anti-$K=8$ state}

We now turn to the particle-hole duals of the above four states. In Appendix C we argue that the anti-$K=8$ state does not exhibit universal behavior in the tunneling experiment with a broad range of possible values for the tunneling exponent $g$. Moreover, we do not expect the quantized conductance $5e^2/2h$ in that state at sufficiently low temperatures and voltages. Its physics is thus quite different from the physics of the other seven states considered in this paper.

\subsection{Anti-331 state}

The edge theory of the anti-331 state can be constructed by reversing the direction of the two FQH edge modes and adding another integer edge channel in the opposite direction to the reversed fractional channels.
The presence of impurities ensures electron tunneling between the fractional edge and the additional integer edge. At weak interaction this tunneling turns out to be irrelevant and the physics become nonuniversal just like in the anti-$K=8$ state. In contrast to the anti-$K=8$ case, the anti-331 state exhibits universal conductance and tunneling exponents at stronger interactions. This is the situation addressed in Appendix A. The FQH edge theory possesses 
the $SO(4)$ symmetry and contains a bosonic charge mode and four upstream Majorana fermions with identical velocities. The Lagrangian density is

\begin{align}
\label{dima12}
\mathcal{L}=&-\frac{2}{4\pi} [\partial_{t}\phi_{\rho}\partial_{x}\phi_{\rho}+v_{\rho} (\partial_{x}\phi_{\rho})^{2}]+i\sum_{k=1}^4\tilde{\psi}_k (\partial_t-\bar{v}_n \partial_x)\tilde{\psi}_k \nonumber\\
& +\frac{\pi }{6}v_{n_1n_2}\varepsilon^{ijkl}\tilde{\psi}_i\tilde{\psi}_j\tilde{\psi}_k\tilde{\psi}_l+\mathcal{L}_2+\mathcal{L}_{\rm int},
\end{align}
where $\tilde\psi_k$ are four Majoranas, the  charge density on the FQH edge is $e\partial_x\phi_\rho/2\pi$ and the contributions $\mathcal{L}_2$ and $\mathcal{L}_{\rm int}$ describe two integer channels as above.

The most relevant quasiparticles carry the electric charge $e/2$. There are four most relevant $e/4$-particle annihilation operators of the structure $\sigma_\alpha(x)\exp(i\phi_\rho/2)$ ($\alpha=1,\dots,4$),
where $\sigma_\alpha$ changes the boundary conditions from periodic to antiperiodic for all four Majorana fermions (see Appendix A for an explicit expression) and has the scaling dimension 1/4.

\subsection{Anti-Pfaffian state}

The anti-Pfaffian state \cite{3,4} is obtained from the Pfaffian state in exactly the same way as the anti-331 state can be obtained from the 331 state. The edge theory is very similar to (\ref{dima12}) and exhibits the $SO(3)$
symmetry. The edge Lagrangian density is

\begin{align}
\label{dima13}
\mathcal{L}=-\frac{2}{4\pi}[\partial_t\phi_\rho\partial_x\phi_\rho+v_\rho(\partial_x\phi_\rho)^2]\nonumber\\
+\sum_{k=1}^3 i\psi_k(\partial_t-\bar{v}_n \partial_x)\psi_k+\mathcal{L}_2+\mathcal{L}_{\rm int}.
\end{align}
The two most relevant $e/4$-quasiparticle operators express as $\sigma_\alpha\exp(i\phi_\rho/2)$ ($\alpha=1,2$), where $\sigma_\alpha$ changes the boundary conditions from periodic to antiperiodic for all Majorana fermions and has the scaling dimension 3/16. 

\subsection{Anti-$SU(2)_2$ state}

The edge theory is similar to the anti-Pfaffian and anti-331 cases. Its derivation is given in Appendix B. There are 5 Majorana fermions on the edge. This corresponds to the emergent $SO(5)$ symmetry. The edge Lagrangian density

\begin{align}
\label{dima14}
\mathcal{L}=-\frac{2}{4\pi}[\partial_t\phi_\rho\partial_x\phi_\rho+v_\rho(\partial_x\phi_\rho)^2]\nonumber\\
+\sum_{k=1}^5 i\psi_k(\partial_t-\bar{v}_n \partial_x)\psi_k+\mathcal{L}_2+\mathcal{L}_{\rm int}
\end{align}
differs from Eq. (\ref{dima13}) only by the number of upstream neutral modes. The four most relevant $e/4$-anyon operators can be written as $\sigma_\alpha\exp(i\phi_\rho/2)$ ($\alpha=1,\dots,4$), where $\sigma_\alpha$ changes the boundary conditions from periodic to antiperiodic for all Majorana fermions and has the scaling dimension 5/16.

\section{Qualitative picture}

In Luttinger liquid systems with position-independent interaction constants, such as the edge theories discussed in Section II, the low-energy tunneling density of states scales as $\rho(E)\sim E^{g-1}$, where $g$ depends on the details of the system \cite{fg}. We argue below that $g$ assumes different values in different geometries. The tunneling conductance is proportional to the product of the tunneling densities of states $\rho_u$ and $\rho_l$ on the upper and lower edges at $E\sim k_B T$  (Ref. \onlinecite{fg}).  In general, the edges are described by  two different exponents $g_u$ and $g_l$.  At low temperatures, the  linear conductance for the tunneling current $G\sim T^{g_u+g_l-2}$. Hence, the experimentally measured $g=(g_u+g_l)/2$.

As we discussed in Section II, in all three geometries the tunneling current is carried by quasiparticles with charge $e/4$. The best fits \cite{20,21,22} for the tunneling exponent at this quasiparticle charge are $g_A=0.45$, $g_B=0.42$ and 
$g_C=0.38$. We want to connect these numbers with the nature of the topological order.

Since the electric current is carried by the edges, the exponent $g$ depends on the edge physics near the QPC. The size of the relevant region near the point contact is set by the distance a charge can travel during the tunneling event: $L=vt$, where $v$ is the velocity of the edge excitation and $t$ is the duration of the tunneling event. The latter can be estimated from the uncertainty relations, $t\sim\hbar/E$, where $E\sim {\rm max} (k_BT,eV)$ is the uncertainty of the quasiparticle energy. Assuming that $v\sim 10^7$ cm/s,  we find $L\sim 10$ $\mu$m at relevant values of the temperature and voltage bias. 

Let us look at the geometry of the edges within  $10$ $\mu$m from the QPC. The edges are defined by gates at the distance of 200 nm from 2DEG. The widths of both gates in the geometry 1a and the upper gate in the geometry 1b have the same order of magnitude of hundreds nanometers \cite{22}. The width is 2200 nm for the lower gate in the geometry 1b and both gates in the geometry 1c. The distance between the edge channel segments on the opposite sides of a gate differs from its width. In our case there are several edge channels that can run at different locations. It is not easy to determine those locations. In particular, the annealing procedure used in Refs. 
\onlinecite{20,21}
changes the device electrostatics but it is unclear what its effect on the geometry is. A crude estimate of the edge channel positions can be obtained from Ref. \onlinecite{csg}. We find that the edge channels run within the distance of hundreds nanometers from the gates. This agrees with the upper bound one can derive from the gate geometry in Fig. 1b. The tip of the upper gate is at 600 nm from the lower gate. The distance between the outermost edge channel and the gate is thus expected to be below 600/2 nm $=300$ nm. We conclude that the gate width, the gate distance from 2DEG and the distance between the edge modes on the opposite sides of the gate all have the same order of magnitude for both gates in the geometry 1a and the upper gate in the geometry 1b. This means that there is a significant unscreened Coulomb interaction between the segments of the edge on the opposite sides of the gates. On the other hand, in the geometry 1c and for the lower gate in the geometry 1b, the width of the gates is close to the distance between the edge channels on their opposite sides and much greater than the gate distance from the 2DEG plane. Thus, we expect that the Coulomb interaction across the gates is almost completely screened in the latter geometries. One can also neglect the electrostatic interaction between the edge channels defined by two different gates in all geometries. Indeed, an edge point at the distance 
$\sim L\sim 10$ $\mu$m from the tip of the gate is much further from the edge on the other side of the QPC than from the gate.

\begin{figure}
\centering
\includegraphics[width=0.7in]{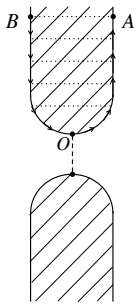}
\caption{Charge tunnels to point O and travels along the edge to point A. Dotted lines show Coulomb repulsion. Arrows show the transport direction on the chiral edge.}
\end{figure}

What is the effect of the electrostatic interaction across the gate on the tunneling current? It is easy to see that repulsive Coulomb interaction suppresses tunneling. Indeed, after a tunneling event, the tunneling charge must move away from the QPC. Due to the chiral transport on the quantum Hall edge, it moves along segment OA in Fig. 2. When the excess charge arrives at point A, it pushes charge from point B due to their repulsive electrostatic interaction. Because of chirality the charge from point B can only move towards the tip O of the gate. Thus, excess charge accumulates in point O. This means lower tunneling density of states than
for a noninteracting system where charge rapidly distributes over a large region.
 Hence, one expects higher values of the tunneling exponent $g$ in the geometries 1a and 1b than in the geometry 1c. This is exactly what has been observed. 

The lower edge has the same geometry in Figs. 1b and 1c and its tunneling density of states is described by the same exponent $g_l$. The same exponent also describes the upper edge in Fig. 1c. Thus, $g_C=g_l$. The upper edge in the geometry 1b is described by a different exponent $g_u$. Hence, $g_B=(g_l+g_u)/2$. If two identical gates of the same geometry as the upper gate in Fig. 1b were used in Fig. 1a then one would obtain $g_A=g_u$. In such situation $g_A$, $g_B$ and $g_C$ form an arithmetic series. Interestingly, this agrees with experiment. At the same time, device 1a was made in a different sample and its geometry details are different from the upper gate in Fig. 1b. Besides, the gate width changes linearly \cite{22} with the distance $x$ from the QPC at the distances  $x\sim L$ in the geometry 1a. This means that the effective interaction strength depends on the temperature since it depends on the gate width at the distance $x\sim \hbar v/k_BT$. Hence, there are corrections to the power law for the tunneling density of states in the geometry 1a. This may be a reason for a poor fit for the quasiparticle charge $e/4$ from the data in that geometry.

\begin{figure}
\centering
\includegraphics[width=1.7in]{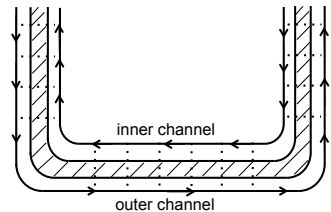}
\caption{The $\Pi$-shaped gate defines an inner and outer quantum Hall channels. Dotted lines illustrate Coulomb interaction across the gate.}
\end{figure}

The above discussion has focused on the geometry 1a and the upper gate in Fig. 1b. We now briefly discuss some features of the lower gate in the geometry 1b and both gates in the geometry 1c. Those gates are $\Pi$-shaped, as shown in Fig. 3, with the width of the metal strips on the order of 200nm. Thus, each gate gives rise to two FQH edge channels inside and outside the gate. We are interested in the tunneling in the outer channel. Its
Coulomb interaction with the inner channel affects the tunneling exponent $g$. We investigate that effect in Section IV.D below and show that it can be neglected, provided that the interaction across the gate is such that the system is not close to the interaction-driven instability.

Our main conclusion is that only $g_C$ is unaffected by the electrostatic interaction between different edge segments. Thus, the information about the topological order at the filling factor 5/2 can be obtained from the comparison of $g_C=0.38$ with theoretical predictions for the models from Section II.  Table I shows the predictions for $g$. We find that the 331 state gives the best fit with an excellent agreement between the theoretical $g=3/8$ and the experimental value $0.38$. As discussed in Section IV.D, for a system on the verge of instability, all three exponents $g_A$, $g_B$ and $g_C$ considerably exceed the universal value in the absence of interactions across the gates.
In this unlikely scenario, the data \cite{21} do not exclude the Pfaffian state.

 \begin {table}
\begin{center}
    \begin{tabular}{ | c | c | c | c | c | p{0.8cm} | p{1.2cm} | p{1.2cm} |}
    \hline
    state & $K=8$ & 331 & Pfaffian&$SU(2)_2$&anti-331&anti-Pfaffian&anti-$SU(2)_2$ \\ \hline
    $g$ & 1/8 & 3/8 & 1/4&1/2&5/8&1/2&3/4 \\ \hline
    \end{tabular}
\end{center}
\caption {Exponent $g$ in the tunneling density of states $\rho(E)\sim E^{g-1}$ for a straight edge in various 5/2 states. 
The values for the $K=8$, 331, Pfaffian, $SU(2)_2$ and anti-Pfaffian states can be found in Ref. \onlinecite{Overbosch}. The anti-331 state is addressed in Appendix A. Appendix B contains calculations for the anti-$SU(2)_2$ state.}
\label{tableI}
\end {table}

\section{Edge Physics and Tunneling Current}

In this section we compute the tunneling exponents $g$ for various models of the 5/2 state in the presence of unscreened Coulomb interaction across the gate. This will allow us to estimate the strength of the electrostatic interaction across the gate. Even though the difference of the tunneling exponents is rather small in the geometries 1b and 1c, we find that the electrostatic interaction is strong: the interaction between the nearest points of the edge segments on the opposite sides of the upper gate in Fig. 1b
turns out comparable with the interaction between the charges, placed in those points in the absence of a gate. This agrees with what one expects from the geometry of the gates.

\subsection{Lagrangian}

The general structure of the action of a straight edge is

\begin{align}
\label{dima15}
L_0=\int dxdt \mathcal{L}=-\frac{1}{4\pi}\int_{-\infty}^{+\infty} dxdt\nonumber\\
 \sum_{ij}[K_{ij}\partial_t\phi_i\partial_x\phi_j
+V_{ij}\partial_x\phi_i\partial_x\phi_j]+ L_{\psi},
\end{align}
where $\phi_i$ denote various edge Bose-modes and $L_{\psi}$ is the action of the Majorana fermion degrees of freedom in the Pfaffian, $SU(2)_2$, anti-$SU(2)_2$, anti-331 and anti-Pfaffian states.
Modes $\phi_i$ include two fields describing integer quantum Hall channels. Stability requires that the symmetric matrix $V_{ij}$ is positive definite.
As seen from Section II, there are no relevant or marginal interactions between the Bose and Majorana degrees of freedom.

\begin{figure}
\centering
\includegraphics[width=2.2in]{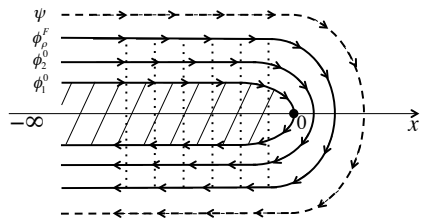}
\caption{Two integer modes $\phi_1^0$ and $\phi_2^0$, the charged FQH mode $\phi^F_\rho$ and the neutral mode $\psi$ travel around the shaded gate. Electrostatic interaction across the gate is illustrated by dotted lines.}
\end{figure}

Consider now the geometry of the edge, defined by a narrow gate, Fig. 4. We can still extend integration in Eq. (\ref{dima15}) from minus to plus infinity, assuming that $x$ is measured along the edge, i.e., 
$x$ is negative on the edge segment above the gate and positive below the gate, Fig. 4. However, Eq. (\ref{dima15}) is no longer the only contribution to the action.
We must  also include the electrostatic interaction across the gate. In the presence of the gate the interaction is short range and hence can be described by local operators coupling fields at the points $x$ and $-x$. Since we consider low temperature physics, we are interested in relevant and marginal operators only. In the following discussion we will assume that the list of such operators remains the same as in the limit of the weak Coulomb interaction. This will allow us to use the same scaling dimensions as in the absence of the electrostatic force across the gate.

The most obvious interaction contribution to the action is $-\frac{1}{4\pi}\int dt \int_{-\infty}^{0}dx\sum_{ij}U_{ij}\partial_x\phi_i(x)\partial_x\phi_j(-x)$.  All other contributions made of derivatives of Bose fields are irrelevant. What about contributions with operators of the form $\exp(i\sum_n a_n\phi_n)$ which shift charge between different edge channels on each side of the gate  (certainly, no charge tunneling across the gate is allowed)? One might think that
some such operators must be included. Indeed, consider the operator $O_1=\exp(i[\phi_1^0(x)+\phi_1^0(-x)-\phi_2^0(x)-\phi_2^0(-x)])$ which moves one electron between the two integer modes on each side of the gate. It appears to be marginal. However, there is inevitable disorder in our system. Disorder can be described by adding contributions of the form $\sum_k\zeta_k(x)\partial_x\phi_k(x)$ with random $\zeta_k(x)$ to the Lagrangian density.
We omitted all such contributions in Eq. (\ref{dima15}) since one can gauge out disorder by redefining $\phi_k\rightarrow\phi_k+2\pi \sum_k(V^{-1})_{ki}\int  \zeta_i dx$. We expect that disorder contains a contribution with no correlation between the opposite sides of the gate. Then our redefinition of the fields introduces randomness in the operator $O_1$ and makes it irrelevant. A similar argument excludes operators, containing $\exp(\pm i\phi_n)$ in the $SU(2)_2$
state. What about other neutral-mode contributions such as operators with Majorana fermions?
The list of possible operators is restricted by the requirement that the topological charge conserves on each side of the gate: Indeed, topological charge cannot travel under the gate where there is no FQH liquid. One might still construct some marginal operators that conserve topological charge on both sides in  some of the states from Section II. For example, the operator $O_2=\tilde\psi_1(x)\tilde\psi_2(x)\tilde\psi_3(-x)\tilde\psi_4(-x)$ is marginal in the anti-331 state. However, all such operators are prohibited by symmetry. Indeed, as discussed in Appendix A, disorder gives rise to the emergent $SO(4)$ symmetry in the anti-331 state. 
The electrostatic coupling between the sides of the gate has the same form as for an infinitely long gate. In the latter case, the system is invariant with respect to two independent $SO(4)$ symmetry groups acting on the two electrostatically coupled edges. 
The combined $SO(4)\times SO(4)$ symmetry excludes the operator $O_2$. A similar argument applies in the anti-Pfaffian and anti-$SU(2)_2$ states.
We conclude that the action can be chosen in the form

\begin{equation}
\label{dima16}
L=L_0-\frac{1}{4\pi}\int dt \int_{-\infty}^{0}dx\sum_{ij}U_{ij}\partial_x\phi_i(x)\partial_x\phi_j(-x).
\end{equation}

\subsection{Tunneling exponent $g$}

The tunneling at the QPC at $x=0$ is described by the contribution to the action

\begin{equation}
\label{dima31}
L_T=\sum_{\alpha,\beta}\int dt \Gamma_{\alpha\beta}\Psi^\dagger_{u,\alpha}(x=0)\Psi_{l,\beta}(x=0)+{\rm H.c.},
\end{equation}
where $\Psi_{u,\alpha}$ and $\Psi_{l,\beta}$ are quasiparticle operators on the upper and lower edges, $\alpha$ and $\beta$ distinguish different quasiparticle operators, $\Gamma_{\alpha\beta}=\Gamma_{\alpha\beta}(E_c)$ are the  tunneling amplitudes
and $E_c$ is the ultraviolet cut-off energy. Certainly, every allowed tunneling process conserves the electric and topological charges. In order to find the low-temperature conductance, it is convenient to perform the renormalization group procedure. We integrate out fast degrees of freedom and rescale $\Gamma_{\alpha\beta}$. Only the terms with the most relevant $e/4$-excitation operators $\Psi_{u,\alpha}^\dagger$ and $\Psi_{l,\beta}$ with the scaling dimensions $\Delta_u$ and $\Delta_l$ must be kept. We stop at the energy scale $E\sim k_B T$. At that scale $\Gamma_{\alpha\beta}(E=k_B T)\sim(k_B T/E_c)^{\Delta_u+\Delta_l-1}\Gamma_{\alpha\beta}(E_c)$. The linear conductance $G\sim T^{2g-2}=T^{g_u+g_l-2}$ can now be found from the perturbation theory: $G\sim|\Gamma(k_BT)|^2$. We conclude that
$g_u=2\Delta_u$, $g_l=2\Delta_l$ and $g=\Delta_u+\Delta_l$.

Thus, to find $g$ we need to compute the scaling dimensions of quasiparticle operators of a general structure $\Psi_{e/4}=\sigma(x=0)\exp(i\sum a_k\phi_k(x=0))$, where the operator $\sigma$ acts on the Majorana sector (and is just an identity operator in some models) and $a_k$ are constants (see Section II). Since the Majorana sector is decoupled from the Bose modes, the scaling dimension of the operator $\sigma$ is not affected by the interaction across the gate and can be taken from Section II. We thus focus on the exponential factor in $\Psi_{e/4}$. 

In all models of Section II, the symmetric $K$-matrix $K_{ij}$ is positive definite.
In the rest of this subsection we also assume that $U_{ij}$, Eq. (\ref{dima16}), is symmetric. This is automatically satisfied, if the gate configuration has a symmetry plane such that reflections in the plane transform the segments of the edge above and below the gate into each other. The gate configuration \cite{20}, Fig. 1a, has such a symmetry plane. The configuration 
of the upper gate,
Fig. 1b, used in Ref. \onlinecite{21}, does not have a symmetry plane but is approximately symmetric beyond about 10 $\mu$m from the QPC. Note that $U_{ij}$ can be symmetric even in the absence of a symmetry plane. This happens if one can neglect all interactions across the gate except the interaction of the charged modes. Then $U_{ij}$ becomes a $1\times 1$ matrix. This is the case in the model of the next subsection. Thus, in Subsection IV.C, we can rely on Subsection IV.B even without assuming the existence of a symmetry plane. Certainly, our qualitative discussion in Section III is also free from that assumption. Calculations are similar but more cumbersome without the symmetry. We focus on the symmetric situation because it allows a proof of two general theorems: 1) in Appendix D we demonstrate that $U_{ij}$ is positive definite; 2) we  show that the positive definite $U_{ij}$ drives the tunneling exponent above the universal value without interaction across the gate. The physical origin of the latter claim has been addressed in Section III.

At this point it is convenient to change notation. We define $\bar\phi_i(x)=(\phi_i(x)+\phi_i(-x))/\sqrt{2}$ and $\theta_i=(\phi_i(x)-\phi_i(-x))/\sqrt{2}$, $x<0$. The Bose-mode contribution to the action now reads

\begin{align}
\label{dima17}
L_B=-\frac{1}{4\pi}\int dt \int_{-\infty}^0 dx\sum_{ij}
 \{
2K_{ij}\partial_t\bar\phi_i\partial_x\theta_j\nonumber\\
+[V_{ij}+\frac{U_{ij}}{2}]\partial_x\theta_i\partial_x\theta_j+[V_{ij}-\frac{U_{ij}}{2}]\partial_x\bar\phi_i\partial_x\bar\phi_j
\}.
\end{align}
The quasiparticle operator becomes $\Psi_{e/4}=\sigma\exp(i\sum a_k\bar\phi_k(0)/\sqrt{2})$. Stability implies that $V\pm U/2$ are positive definite matrices.
We next diagonalize the bilinear form $K_{ij}\partial_t\bar\phi_i\partial_x\theta_j$ with the transformation
$\bar\phi_i=\sum_k S_{ik}\tilde\phi_k;\theta_i=\sum_k S_{ik}\tilde\theta_k$, where the matrix $S$ is such that $S^T K S=E$. The Bose contribution to the action becomes

\begin{align}
\label{dima18}
L_B=-\frac{1}{4\pi}\int dt \int_{-\infty}^0 dx \sum_{ij}
\{
2\delta_{ij}\partial_t\tilde\phi_i\partial_x\tilde\theta_j\nonumber\\
+[\tilde V_{ij}+\tilde\Lambda_{ij}]\partial_x\tilde\theta_i \partial_x\tilde\theta_j
+[\tilde V_{ij}-\tilde\Lambda_{ij}]\partial_x\tilde\phi_i \partial_x\tilde\phi_j
\},
\end{align}
where $\tilde V=S^T V S$ and $\tilde\Lambda=S^T U S/2$ are positive definite symmetric matrices. Integrating out the fields $\tilde\theta_i$ one gets

\begin{align}
\label{dima19}
L_B=\frac{1}{4\pi}\int dt \int_{-\infty}^0 dx\sum_{ij} 
\{
(\tilde V+\tilde\Lambda)^{-1}_{ij}\partial_t\tilde\phi_i\partial_t\tilde\phi_j\nonumber\\
-(\tilde V-\tilde\Lambda)_{ij}\partial_x\tilde\phi_i\partial_x\tilde\phi_j
\}.
\end{align}
The quasiparticle operator assumes the form $\Psi_{e/4}=\sigma\exp(i\sum_{kj} a_k S_{kj} \tilde\phi_j(0)/\sqrt{2})$. 

We introduce another piece of notation now. Consider a symmetric positive definite matrix $A$. We define $\sqrt{A}$ in the following way. We first find such orthogonal matrix $O$ that $A=O^T \bar A O$, where
$\bar A$ is a diagonal matrix with positive diagonal entries. Next, we define $\sqrt{\bar A}$ as a diagonal matrix with positive diagonal entries such that $(\sqrt{\bar A})^2=\bar A$. Finally, we set $\sqrt{A}=O^T \sqrt{\bar A} O$.
Obviously, $\sqrt{A}$ is a symmetric positive definite matrix whose square is $A$; $(\sqrt{A})^{-1}=\sqrt{A^{-1}}$. We make the following change of variables

\begin{equation}
\label{dima20}
\tilde\phi_i=\sum_j\left(\sqrt{\tilde V+\tilde\Lambda}\left[{\sqrt[4]{\sqrt{\tilde V+\tilde\Lambda}(\tilde V-\tilde \Lambda)\sqrt{\tilde V+\tilde \Lambda}}}\right ]^{-1}\right)_{ij}\hat\phi_j.
\end{equation}
The action becomes

\begin{equation}
\label{dima21}
L_B=\frac{1}{4\pi}\int dt \int_{-\infty}^0 dx \sum_{ij}\left\{
P^{-1}_{ij}\partial_t\hat\phi_i\partial_t\hat\phi_j-P_{ij}\partial_x\hat\phi_i\partial_x\hat\phi_j,
\right\}
\end{equation}
where

\begin{equation}
\label{dima22}
P=\sqrt{\sqrt{\tilde V+\tilde\Lambda}(\tilde V-\tilde \Lambda)\sqrt{\tilde V+\tilde \Lambda}}.
\end{equation}

At this point we trace back the steps that led us to Eq. (\ref{dima19}) with Eq. (\ref{dima21}) as the starting point. 
We introduce auxiliary fields $\hat{\theta}_i$ and rewrite the action in the form, similar to Eq. (\ref{dima18}):
\begin{align}
\label{insert1}
L_B=-\frac{1}{4 \pi}\int dt \int_{-\infty}^{0}dx \sum_{ij}\{2\delta_{ij}\partial_t\hat{\phi}_i \partial_x\hat{\theta}_j \nonumber \\
+P_{ij}\partial_x\hat{\theta}_i \partial_x\hat{\theta}_j +P_{ij}\partial_x\hat{\phi}_i \partial_x\hat{\phi}_j\}.
\end{align}
Next, we define new fields $\Phi_i$ according to $\hat{\phi}_i(x)=(\Phi_i(x)+\Phi_i(-x))/\sqrt{2}$ and $\hat{\theta}_i(x)=(\Phi_i(x)-\Phi_i(-x))/\sqrt{2}$, $x<0$, and end up with the action

\begin{equation}
\label{dima23}
L_B=-\frac{1}{4\pi}\int_{-\infty}^{+\infty}dt dx \sum_{ij}\left[\delta_{ij}\partial_t\Phi_i\partial_x\Phi_j+P_{ij}\partial_x\Phi_i\partial_x\Phi_j\right],
\end{equation}
In Eq. (\ref{dima23}) we return to our initial definition of the coordinate
$-\infty<x<+\infty$ and the points $x$ and $-x$ no longer interact. The quasiparticle operator $\Psi_{e/4}=\sigma\exp\left[i\sum_{kj}a_k\left(S\sqrt{\tilde V+\tilde\Lambda}\sqrt{P^{-1}}\right)_{kj}\Phi_j(0)\right]$.
It is now easy to write down the scaling dimension $\Delta$ of $\Psi_{e/4}$ and $g=2\Delta$:

\begin{equation}
\label{dima24}
g=2\Delta_\sigma+\sum_{ij}\left(S\sqrt{\tilde V+\tilde\Lambda}P^{-1}\sqrt{\tilde V+\tilde\Lambda}S^T\right)_{ij}a_i a_j,
\end{equation}
where $\Delta_\sigma$ is the scaling dimension of the operator $\sigma$.

We now prove that for any choice of $a_i$, i.e., for any quasiparticle operator, $g$ in Eq. (\ref{dima24})
exceeds the value one would obtain without Coulomb interaction between the edge segments on the opposite sides of the gate. 
In the above expression the electrostatic interaction across the gate enters only through the symmetric positive definite matrix $\tilde\Lambda$. 
We thus wish to prove that $g$, Eq. (\ref{dima24}), is greater at $\tilde\Lambda\ne 0$ than at $\tilde\Lambda=0$. 

$\Delta_\sigma$ does not depend on $\tilde\Lambda$. The symmetric positive definite matrix $N=\sqrt{\tilde V+\tilde\Lambda}P^{-1}\sqrt{\tilde V+\tilde\Lambda}$ that enters (\ref{dima24}) reduces to the identity matrix at $\tilde\Lambda=0$. Hence, it is enough to prove that all eigenvalues $n_i$ of $N$ are greater than 1. For this end, we notice that

\begin{equation}
\label{dimaE1}
N^{-1}(\tilde V+\tilde\Lambda)N^{-1}=(\tilde V-\tilde\Lambda).
\end{equation}
Let us work in the eigenbasis of $N$ and look at the diagonal elements of the above Eq. (\ref{dimaE1}). We find

\begin{equation}
\label{dimaE2}
\frac{1}{n_i^2}{\left(\tilde V_{ii} +\tilde \Lambda_{ii}\right)}=\left(\tilde V_{ii} -\tilde\Lambda_{ii}\right)
\end{equation}
and hence 
$n_i=+1/\sqrt{1-\frac{2\tilde\Lambda_{ii}}{\tilde V_{ii}+\tilde\Lambda_{ii}}}$.
Since both matrices $\tilde\Lambda$ and $(\tilde V+\tilde\Lambda)$ are positive definite, their diagonal matrix elements are positive. Hence, $n_i>1$ and $g$ (\ref{dima24}) increases indeed at nonzero $\tilde\Lambda$ compared to the case of $\tilde\Lambda=0$. This agrees with the relation between the exponents $g_A$, $g_B$ and $g_C$ in Refs. \onlinecite{20,21}.

\subsection{Estimates of the electrostatic force}

In this subsection we address an apparent paradox: The geometry suggests a rather strong interaction across the upper gate in geometry 1b; why is then the tunneling exponent $g_B$ close to $g_C$?

The expression (\ref{dima24}) depends on several unknown parameters. They cannot be extracted from a single observable $g$. In particular, one cannot compute the strength of the interaction across the gate. We can only estimate its order of magnitude from $g_C$ and $g_B$. Our estimates show that despite a small difference $g_B-g_C$, the interaction is not weak. Interestingly, our estimates do not depend on the nature of the bulk topological order.

The estimates will be based on two simple models. In the first model we will assume that the interaction contribution $U$ to the action reduces to the product of the charged modes on the opposite sides of the gate,

\begin{equation}
\label{dima25}
U=-\frac{2}{4\pi}\int dt\int_{-\infty}^0 dx \frac{2}{5}\lambda_1\partial_x\Phi_\rho(x)\partial_x\Phi_\rho(-x),
\end{equation}
where $e\partial_x\Phi_\rho/2\pi$ stays for the linear charge density. Such model is legitimate, if all edge channels on each side of the gate run much closer to each other than to the gate.
In our second model  the across-the-gate interaction includes only the charge density in the FQH channel:

\begin{equation}
\label{dima26}
U=-\frac{2}{4\pi}\int dt\int_{-\infty}^0 dx 2 \lambda_2\partial_x\Phi^F_\rho(x)\partial_x\Phi^F_\rho(-x),
\end{equation}
where $e\partial_x\Phi^F_\rho/2\pi$ is the linear charge density on the edge channel, separating the $\nu=2$ and $\nu=5/2$ regions. 
Model 2 is legitimate, if the gate is close to 2DEG and all integer channels run under the gate while fractional channels are sufficiently far from the gate.

It is unlikely that either of the above two sets of assumptions accurately describes the experimental system. At the same time, our estimates of $\lambda$ in Eqs. (\ref{dima25}) and (\ref{dima26}) will give an idea of the range of possible interaction strength.

For simplicity we will assume that the mode $\Phi_\rho$ in model 1 and the mode $\Phi^F_{\rho}$
in model 2 do not interact with any other edge modes on the same side of the gate. The application of Eq. (\ref{dima24}) then becomes very easy. 

Consider first model 1. We can always make a linear change of the variables $\phi_i$ such that one of the new variables is $\Phi_\rho$ and all other new variables commute with $\Phi_{\rho}$. Then the dynamics of the charged mode is completely independent from all other modes. The form of the charged mode action is dictated \cite{wen} by the quantum Hall conductance $\frac{5e^2}{2h}$:

\begin{equation}
\label{dima27}
L_{\rho}=-\frac{1}{4\pi}\int dt \int_{-\infty}^{+\infty} dx \left[\frac{2}{5}\partial_t\Phi_\rho\partial_x\Phi_\rho+\frac{2}{5}v_{\rho}\partial_x\Phi_\rho\partial_x\Phi_\rho\right],
\end{equation}
where $v_\rho$ is the velocity of the mode.
Since the quasiparticle charge is $e/4$, the field $\Phi_\rho$ must enter the quasiparticle operator as $\exp(i\Phi_\rho/[4\times\frac{5}{2}])=\exp(i\Phi_\rho/10)$. Finally, from Eq. (\ref{dima24}) one finds:

\begin{equation}
\label{dima28}
g_B-g_C=\frac{1}{80}\left[\sqrt{\frac{v_\rho+\lambda_1}{v_\rho-\lambda_1}}-1\right].
\end{equation}

In model 2 we similarly assume that the charged mode of the FQH edge is decoupled from all other modes. It is controlled by the action

\begin{equation}
\label{dima29}
L_{\rho}=-\frac{1}{4\pi}\int dt \int_{-\infty}^{+\infty} dx \left[2\partial_t\Phi^F_\rho\partial_x\Phi^F_\rho+2v_{\rho}\partial_x\Phi^F_\rho\partial_x\Phi^F_\rho\right],
\end{equation}
where the coefficient 2 reflects \cite{wen} the FQH conductance $\frac{e^2}{2h}$. The quasiparticle operator contains $\Phi^F_\rho$ in the exponential factor $\exp(i\Phi^F_\rho/[4\times\frac{1}{2}])=\exp(i\Phi^F_{\rho}/2)$. Hence, Eq. (\ref{dima24})
yields

\begin{equation}
\label{dima30}
 g_B-g_C=\Delta g= \frac{1}{16}{\left[\sqrt{\frac{v_\rho+\lambda_2}{v_\rho-\lambda_2}}-1\right]}.
\end{equation}

Substituting the experimental $g_B-g_C=0.04$ in Eqs. (\ref{dima28},\ref{dima30}) we find $\lambda_1/v_\rho\sim 0.9$ and $\lambda_2/v_\rho\sim 0.45$.
These values are similar to the estimate \cite{41} in a related geometry in the 1/3 state.
Certainly, the above two models are not very realistic. Besides, even small uncertainties in $g_B$ and $g_C$ result in major uncertainties in $\lambda_{1,2}$.
Note also that we neglect the repulsive interaction between the upper and lower edges of the 2200 nm-wide constriction in the geometry 1c since the constriction width is much shorter than the relevant thermal length. This interaction decreases $g_C$ and hence decreases the estimates (\ref{dima28},\ref{dima30}) for $\lambda_{1,2}$.
Finally, in the above discussion we disregarded the repulsive interaction between the edge modes on the inner and outer sides of the gates in the geometry 1c and the lower gate in the geometry 1b (Fig. 3). 
This interaction increases both $g_B$ and $g_C$ compared to the situation without an inner edge.
In the next subsection we show that it is safe to neglect the interaction of the inner and outer channels unless it is so strong that the system is on the verge of instability. This possibility is not a concern for us here because our goal is to demonstrate that the interaction is not weak. Overall, it appears  safe to conclude that physical $\lambda/v_\rho$ is within an order of magnitude from the above values. This corresponds to a significant interaction across the gate.

\subsection{Interaction of inner and outer edge channels}

In this subsection we address the interaction between the inner and outer edge channels,  Fig. 3, in the geometry 1c and around the lower gate in the geometry 1b. We find that for most values of parameters this interaction has considerably less effect on the tunneling exponents $g_B$ and $g_C$ than the interaction discussed in the previous subsection. This justifies neglecting the interaction of the inner and outer channels.

To estimate the change in the exponents $g_B$ and $g_C$ we consider a model, similar to the second model of the previous subsection. The action reads

\begin{align}
\label{4D1}
L=-\frac{2}{4\pi}\int dt\int_{-\infty}^{+\infty}dx[\partial_t\Phi^F_{\rho O}\partial_x\Phi^F_{\rho O}+v_\rho\partial_x\Phi^F_{\rho O}\partial_x\Phi^F_{\rho O} \nonumber\\
-\partial_t\Phi^F_{\rho I}\partial_x\Phi^F_{\rho I}+v_\rho\partial_x\Phi^F_{\rho I}\partial_x\Phi^F_{\rho I}+2\lambda_2\partial_x\Phi^F_{\rho O}\partial_x\Phi^F_{\rho I}],
\end{align}
where $\Phi^F_{\rho O}$ and $\Phi^F_{\rho I}$ describe the charge density in the outer and inner fractional edge channels. We are interested in the tunneling into the outer channel. $\Phi^F_{\rho O}$ enters
the tunneling operator through the factor $\exp(i\Phi^F_{\rho O}/2)$. Thus, we wish to compute the correction $\delta g$ to the scaling dimension of the above operator due to a nonzero $\lambda_2$.

It is convenient to change the variables:

\begin{align}
\label{4D2}
\Phi^F_{\rho O}=\frac{(z+1/z)}{2}\phi_+ +\frac{(z-1/z)}{2}\phi_-; \nonumber\\
\Phi^F_{\rho I}=\frac{(z-1/z)}{2}\phi_+  + \frac{(z+1/z)}{2} \phi_ -,
\end{align}
where $z=\sqrt[4]{\frac{v_\rho-\lambda_2}{v_\rho+\lambda_2}}$. In the new variables, we discover the action of two noninteracting chiral fields:

\begin{align}
\label{4D3}
L=-\frac{2}{4\pi}\int dt \int dx \{\partial_t\phi_+\partial_x\phi_+ -\partial_t\phi_-\partial_x\phi_- \nonumber\\
+ \sqrt{v^2_{\rho}-\lambda_2^2}[(\partial_x\phi_+)^2+(\partial_x\phi_-)^2]\}.
\end{align}
Computing $\delta g$ is now easy. One finds

\begin{equation}
\label{4D4}
\delta g=\frac{1}{16}\left(\frac{1}{\sqrt{1-(\lambda_2/v_\rho)^2}}-1\right).
\end{equation}

The experimentally measured $g$ differs from the universal value $g_{\rm universal}$ for a system without an inner channel along a wide gate and Coulomb interaction across a narrow gate: $g_C=g_{\rm universal}+2\delta g$ and 
$g_B=g_{\rm universal}+\delta g +\Delta g$, where $\Delta g$ is given by Eq. (\ref{dima30}) and reflects the effect of the interaction across the upper gate in Fig. 1b. In the previous subsection we estimated
$\lambda_2/v_\rho\sim 0.45$. This corresponds to $\Delta g\sim 0.04$. Substituting the same value of $\lambda_2$ in Eq. (\ref{4D4}), one finds $\delta g\sim 0.007$. This justifies neglecting $\delta g$ above. 
It is easy to see that $\delta g$ is much less than 1 and considerably smaller than
$\Delta g$ as long as $\lambda_2/v_\rho$ is not close to 1. The unlikely regime $\lambda_2/v_\rho\approx 1$ describes a system on the verge of an instability due to Coulomb interaction. Even in that regime $\delta g<\Delta g/2$ for all 
$\lambda_2$ and hence $\delta g$ can be ignored in crude estimates of, e.g., the Coulomb interaction strength. At the same time, the expressions (\ref{dima28}), (\ref{dima30}) and (\ref{4D4}) all diverge as 
$\lambda_{1,2}/v_\rho\rightarrow 1$ and all three experimental exponents $g_A$, $g_B$ and $g_C$ provide only upper bounds for $g_{\rm universal}$ in that unlikely limit. 

One can avoid complications due to the inner channel by changing the geometry of the gates. Instead of  $\Pi$-shaped gates, one can use gates that contain not only the perimeter but also the inside of a rectangle.

\section{Discussion}

\begin{figure}
\centering
\includegraphics[width=1in]{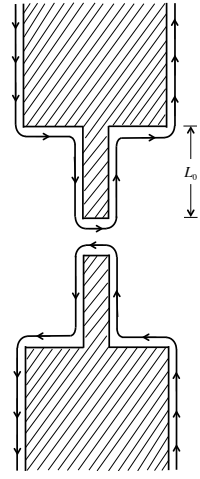}
\caption{Setup of the ``smoking gun'' experiment. The width of the gates depends on the distance from the tunneling contact.}
\end{figure}

We find that the 331 state with the theoretical $g_C=3/8$ gives the best fit to the experimental $g_C=0.38$.
It is instructive to compare this conclusion with the lessons from other types of experiments. The data on the spin polarization are controversial \cite{mstern10,rhone11,tiemann12,mstern12}. 
There is support for both 0 and 100\% polarization. The 331 state comes in both spin-polarized and unpolarized versions with identical transport signatures in the tunneling experiment. Most other proposed 5/2 states are spin-polarized. Some features of the thermopower data \cite{chickering13}
are qualitatively compatible with theoretical predictions for the Pfaffian state \cite{yang09}. However, even Abelian states may exhibit qualitatively similar behavior, if different quasiparticle species are degenerate or close in energy. The Fabry-Perot interference experiments \cite{willett09,willett10} were interpreted \cite{bishara09} as supporting the Pfaffian or anti-Pfaffian state. At the same time, it was shown that the 331 state can produce identical signatures in a Fabry-Perot interferometer \cite{fp331} (but interestingly, not in a Mach-Zehnder interferometer \cite{16}).  The anti-331 state may also exhibit identical signatures.
The results of the transport experiment \cite{36} were explained as a sign of an upstream neutral mode. Such interpretation is incompatible with all proposed states except anti-Pfaffian, anti-331 and anti-$SU(2)_2$.
In particular, the 331 state has no upstream edge modes. As discussed in section III, in the unlikely case of a system on the verge of Coulomb-interaction driven instability, the data \cite{21} do not exclude the Pfaffian state. That state also does not possess contra-propagating edge modes.

Thus, a conflict appears between the interpretation of the experiments from Refs. \onlinecite{36} and \onlinecite{20,21}.  Our explanation of the data \cite{20,21} is based on the assumption that the system is in the scaling regime, where universal predictions apply. We also assume that the edges can be described by a chiral Luttinger liquid model. It may happen that such description fails due to, e.g., dissipation \cite{braggio12}. At the same time, edge reconstruction \cite{cw} or bulk transport \cite{bulk}  may affect the interpretation of the experiments on upstream modes.  On the other hand, various proposed 5/2 states are rather close in energy \cite{biddle} and it was suggested that more than one 5/2 state might be present at different conditions  or in different samples \cite{jain}. It is also possible that the true 5/2 state is not one of the simplest states considered above. More light could be shed on the nature of the 5/2 state by tunneling experiments at lower temperatures. To reduce the effect of Coulomb interaction across the gate, one needs to modify the 1c setup, Fig. 3. The inner edge channel can be eliminated by filling the inside of the $\Pi$-shaped gate, Fig. 3, with metal. Other methods that could show unique signatures of Abelian and non-Abelian states include thermopower measurements \cite{yang09,chickering13}
and Mach-Zehnder interferometry \cite{14,16,mz,13,15,ponomarenko07,ponomarenko10,law08,double}. 
The non-equilibrium fluctuation-dissipation theorem \cite{fdt1,fdt2} would provide an independent  test of the existence of upstream modes.

Our key idea about the role of gating in the experiments \cite{20,21} can be tested with a ``smoking gun'' experiment illustrated in Fig. 5. The width of the gate is different at short distances $x<L_0$ and long distances $x>L_0$ from the gate. The Coulomb interaction across the gate is strong at $x<L_0$ and negligible at large $x$. In such situation our theory predicts two different power dependencies of the conductance on the temperature at $T< \hbar v/k_BL_0$ and $T>\hbar v/k_BL_0$, where $v$ is the excitation speed on the edges. The higher temperature regime is similar to the geometry 1a and the low temperature limit corresponds to the geometry 1c. Thus, one can go between two fixed points by simply changing the temperature (or voltage).

Even though the agreement between the experimental $g_C=0.38$ and theoretical $g_C=3/8$ looks excellent, one should be cautious about data accuracy. For example, electron tunneling experiments into the $1/3$ edge
have routinely shown a 10\% discrepancy with the theory \cite{chang2003}. A relative fragility of the 5/2 edge may lead to even lower experimental resolution. Nevertheless, the difference between $g_C$ in the 331 and other competing states exceeds 30\% in all cases and this lends credibility to the 331 state as the best fit.

A considerable body of numerical work supports Pfaffian or anti-Pfaffian states (see, e.g., Refs. \onlinecite{morf98,RH00,wan06,wan08}). As discussed above, the data \cite{20,21} are not compatible with the anti-Pfaffian state but do not exclude the Pfaffian state in the unlikely case of the system on the verge of instability. On the other hand, numerical studies of small simplified model systems have limitations and only experiment can determine the right state. In particular, the existing numerical results \cite{morf2002,morf2003,feiguin2008,storni2010} for the energy gap at $\nu=5/2$ significantly exceed experimental findings. One limitation of numerics is due to the fact that most studies assume full spin polarization outright. This would exclude the spin-unpolarized 331 state. At the same time, several papers \cite{morf98,dimov08,feiguin2009,biddle} lend support to a spin-polarized ground state. Another limitation comes from the incomplete understanding of the Landau-level mixing effects \cite{wojs2006,wojs2010,rezayi2010,peterson2013}. Disorder  apparently plays a major role in the discrepancy of the theoretical and experimental results for the energy gap \cite{morf2003,dAmbrumenil2011}  and may affect the nature of the ground state. We are not aware of any numerical studies that include disorder. Moreover, the numerical energy difference between 
competing states  \cite{biddle} is so small that even a tiny and generally neglected effect of the spin-orbit interaction \cite{bychkov,dikman} might be relevant.

In conclusion, we have proposed an explanation of the discrepancy of the tunneling exponents in Refs. \onlinecite{20,21} Two of the three geometries used in Refs. \onlinecite{20,21} are affected by the electrostatic interaction across the gates that changes the exponent $g$.
We compare the exponent $g_C$ in the third geometry 1c with the theoretical predictions for seven states with the simplest topological orders at $\nu=5/2$. For some of those states, the tunneling exponents were computed earlier.
In addition, we compute the tunneling exponents for the particle-hole conjugates of the non-Abelian $SU(2)_2$ state and Abelian 331 state in Appendices A and B. The latter state can also be viewed as the Bonderson-Slingerland state built on top of the particle-hole conjugate of the Laughlin 1/3 state. We find that the 331 state gives the best fit to the experiment.

\begin{acknowledgments}
We thank C. Dillard, M. A. Kastner, X. Lin, N. E. Staley and C. Wang for useful discussions. This work was supported by the NSF under Grant No. DMR-1205715.
\end{acknowledgments}

\appendix
\section{Edge of the anti-331 state}
In this appendix we formulate the edge theory of the  particle-hole conjugate of the spin-polarized $331$ state which we call the anti-331 state.
We only consider the FQH edge between the $\nu=2$ and $\nu=5/2$ regions. The two additional integer edge modes will play little role in our discussion.

 We use an approach similar to the treatment of the disorder-dominated $\nu=2/3$ state\cite{38} and the anti-Pfaffian state\cite{3,4}. We start from a clean particle-hole conjugate of the $331$ state which is obtained by condensing hole excitations of the $\nu=1$ quantum Hall state into the $331$ state. In our picture, the 2DEG has an annular shape with the inner edge being the interface between the hole $331$ state and its parental $\nu=1$ IQH state and the outer edge being the interface between the $\nu=1$ quantum Hall state and the vacuum. The edge is the combination of a right-moving $\nu=1$ edge and a left-moving FQH edge of the $331$ state with the Lagrangian density
\begin{equation}
\label{dimaA1}
\mathcal{L}_0= -\frac{1}{4\pi}  \sum_{ I,J=0,1,2}  [K_{IJ}\partial_{t}\phi_{I}\partial_{x}\phi_{J}+V_{IJ}\partial_{x}\phi_{I}\partial_{x}\phi_{J}],
\end{equation}
where the $K$-matrix \cite{wen} and the potential matrix are
\begin{equation}
\label{dimaA2}
K =
\left( \begin{array}{ccc}
1& 0 & 0 \\
0& -3 & 2 \\
0& 2 & -4 \end{array} \right)
\mspace{9.0mu} \textrm{and} \mspace{9.0mu}
V =
\left( \begin{array}{ccc}
V_{00} & V_{01} & V_{02} \\
V_{01} & V_{11} & V_{12} \\
V_{02} & V_{12} & V_{22} \end{array} \right),
\end{equation}
$\phi_{0}$ is the right-moving $\nu=1$ charge density mode, $\phi_{1}$ and $\phi_{2}$ are two left-moving modes in the edge theory of the $331$ state. The charge vector that determines how the  modes ($\phi_0,\phi_1,\phi_2$) couple to the external gauge field and contribute to the electric current is $\mathbf{t}^T=(1,1,0)$. The $K$-matrix and the charge vector $\mathbf{t}$ together characterize the topological order of the FQH state. To simplify the discussion, we assume at this point that $V_{12}=-V_{22}/2$. We will also assume that the $\nu=1$ edge mode only couples with the edge mode in the bottom hierarchy of the hole $331$ state so that $V_{02}=0$. This approximation 
reflects the fact that only the bottom level hierarchy contains electrons which interact with the $\nu=1$ IQH liquid through the Coulomb potential. At the end of the appendix, we will relax the above assumptions and will see that they do not change any conclusions.

The Lagrangian $\mathcal{L}_0$ describes the low energy edge physics of a clean particle-hole conjugate of the $331$ state. In reality, there are always impurities on the edge. They destroy the translational invariance on the FQH edge and cause inter-edge tunneling which leads to the edge mode equilibration. Nontrivial topological charges cannot travel between the hole $331$ edge and the $\nu=1$ channel. Hence, only electrons can tunnel. The electron operator on the $\nu=1$ edge is $e^{i\phi_0}$. The most relevant electron operators on the hole $331$ edge are $e^{-i(3\phi_{1}-2\phi_{2})}$ and $e^{-i(\phi_{1}+2\phi_{2})}$. The appropriate term for the electron tunneling in the Lagrangian density  is
\begin{equation}
\label{dimaA3}
\mathcal{L}_{\textrm{tun}}=\xi_{1}(x) e^{i(\phi_{0}+3\phi_{1}-2\phi_{2})}+\xi_{2}(x) e^{i(\phi_{0}+\phi_{1}+2\phi_{2})}+  \textrm {H.c.},
\end{equation}
where we have suppressed the Klein factors that are necessary to ensure correct statistics among different electron operators. $\xi_{1}(x)$ and $\xi_{2}(x)$ are complex variables characterizing the strength of random impurities. We assume for simplicity that their distribution is Gaussian and they are $\delta$-correlated: $\langle \xi_{1}(x)\xi_{1}^*(x')\rangle =W_1 \delta(x-x')$ and $\langle \xi_{2}(x)\xi_{2}^*(x')\rangle =W_2 \delta(x-x')$. 

At weak disorder $W_{1}$ and $W_2$, the effect of electron tunneling on the FQH edge is determined by the scaling behavior of the tunneling operators. Using $\mathcal{L}_0$, we find that the two tunneling operators have the same scaling dimension
\begin{equation}
\label{dimaA4}
\Delta=\frac{1}{2}+\frac{3-2\sqrt{2}c}{2\sqrt{1-c^2}},
\end{equation}
where $c=8\sqrt{2}V_{01}/(8V_{00}+4V_{11}-V_{22})$. 
The leading order renormalization group (RG) equations for the disorder strength are\cite{39}
\begin{equation}
\label{dimaA5}
\frac{dW_i}{dl}=(3-2\Delta)W_i,
\end{equation}
where $i=1,2$.
For $\Delta>3/2$, the electron tunneling is irrelevant and the low temperature edge physics is described by $\mathcal{L}_0$ in Eq.~(\ref{dimaA1}) in which no equilibration occurs between the $\nu=1$ edge and the hole $331$ edge. 
The physics is very similar to the physics of the anti-$K=8$ state, addressed in Appendix C.
Quasiparticle tunneling is nonuniversal. At low voltages and temperatures the quantum Hall conductance is quantized at $7e^2/2h$ instead of the right value $5e^2/2h$. This happens because the integer mode 
conductance $3e^2/h$ and the conductance $e^2/2h$ of the fractional edge must be added and not subtracted in the quantum Hall bar geometry. A detailed discussion of a similar point can be found in Ref. \onlinecite{wang10a}.
On the other hand, if $\Delta<3/2$ the electron tunneling  is relevant and we end up with a disorder-dominated phase  with equilibrated edge modes at low temperatures. This is the situation we consider below.

To study the disorder-dominated phase, it is useful to rewrite the edge dynamics in terms of a charged mode, represented by the charge vector $\mathbf{t}$, and two independent neutral modes, represented by the neutral vectors $\mathbf{n_{1}}$, $\mathbf{n_{2}}$. If we treat the matrix $K^{-1}$ as a metric then neutral vectors are those with  vanishing inner products with the charge vector, $\mathbf{n}_{1}^{T}  K^{-1} \mathbf{t}=0$, $\mathbf{n}_{2}^{T}  K^{-1} \mathbf{t}=0$. We choose $\mathbf{n}_{1}^{T}=(1,2,0)$, $\mathbf{n}_{2}^{T}=(0,-1,2)$. The corresponding charged $\phi_\rho$ and neutral $\phi_{n_1}$, $\phi_{n_2}$ boson fields are
\begin{eqnarray}
\label{dimaA6}
\mathbf{t} &\rightarrow& \phi_{\rho}=\phi_{0}+\phi_{1} \nonumber \\ 
\mathbf{n}_{1} &\rightarrow& \phi_{n_{1}}= \phi_{0}+2 \phi_{1} \nonumber \\
\mathbf{n}_{2} &\rightarrow& \phi_{n_{2}}= -\phi_{1}+2\phi_{2}.
\end{eqnarray}
In the basis ($\phi_{\rho} ,\phi_{n_1},\phi_{n_2} $), the Lagrangian density of the tunneling problem $\mathcal{L}_0+\mathcal{L}_{\textrm{tun}}=\mathcal{L}_{\rho}+\mathcal{L}_{\textrm{Sym}}+\mathcal{L}_{\textrm{SB}}$, where
\begin{equation}
\label{dimaA7}
\mathcal{L}_{\rho}=-\frac{2}{4\pi} [\partial_{t}\phi_{\rho}\partial_{x}\phi_{\rho}+v_{\rho} (\partial_{x}\phi_{\rho})^{2}]
\end{equation}
and
\begin{align}
\label{dimaA8}
\mathcal{L}_{\textrm{Sym}}=& -\frac{1}{4\pi} \sum_{i=1,2} [-\partial_{t}\phi_{n_{i}}\partial_{x}\phi_{n_{i}}+\bar{v}_{n} (\partial_{x}\phi_{n_{i}})^{2}] \nonumber \\
&+ [\xi_{1}(x) e^{i(\phi_{n_{1}}-\phi_{n_{2}})}+\xi_{2}(x) e^{i(\phi_{n_{1}}+\phi_{n_{2}})}+  \textrm {H.c.}]; \nonumber \\
\mathcal{L}_{\textrm{SB}}=& -\frac{1}{4\pi} \sum_{i=1,2}\delta v_{n_i} (\partial_{x}\phi_{n_{i}})^{2}-\frac{2}{4\pi}v_{\rho n_1}\partial_{x}\phi_{\rho}\partial_{x}\phi_{n_{1}}.
\end{align} 
The velocities of $\phi_{\rho}$, $\phi_{n_1}$ and $\phi_{n_2} $ are $v_{\rho}=2V_{00}-2V_{01}+\frac{1}{2}V_{11}-\frac{1}{8}V_{22}$, $v_{n_1}=V_{00}-2V_{01}+V_{11}-\frac{1}{4}V_{22}$ and $v_{n_2}=\frac{1}{4}V_{22}$, respectively. In writing the Lagrangians, we have defined the average velocity $\bar{v}_{n}=\frac{1}{2}(v_{n_1}+v_{n_2})$ and moved the anisotropic part of the velocities $\delta v_{n_i}= v_{n_i}- \bar{v}_{n}$ into $\mathcal{L}_{\textrm{SB}}$. The reason for this is that $\mathcal{L}_{\textrm{Sym}}$ now has a hidden $SO(4)$ symmetry whereas $\mathcal{L}_{\textrm{SB}}$ is the symmetry-breaking term. The interaction between the charged mode $\phi_{\rho}$ and the neutral mode $\phi_{n_1}$ is $v_{\rho n_1}=-2V_{00}+3V_{01}-V_{11}+\frac{1}{4}V_{22}$. Note that after the basis change the electron tunneling only couples the neutral modes and contributes to their equilibration. The charged mode is left alone with its own velocity $v_{\rho}$. 

To view the $SO(4)$ symmetry in $\mathcal{L}_{\textrm{Sym}}$ more clearly, we fermionize $\phi_{n_1}$ and $\phi_{n_2} $ into chiral fermions by setting $\Psi_{n_1}=\frac{1}{\sqrt{2 \pi}}e^{-i\phi_{n_1}}$ and $\Psi_{n_2}=\frac{1}{\sqrt{2 \pi}}e^{-i\phi_{n_2}}$. We then further break the two chiral fermions into four real Majorana fermions defined as $\psi_1=\textrm{Re}\Psi_{n_1}$, $\psi_2=\textrm{Im}\Psi_{n_1}$, $\psi_3=\textrm{Re}\Psi_{n_2}$ and $\psi_4=\textrm{Im}\Psi_{n_2}$. In terms of the Majorana fermions, the Lagrangians $\mathcal{L}_{\textrm{Sym}}$ and $\mathcal{L}_{\textrm{SB}}$ are
\begin{eqnarray}
\label{dimaA9}
\mathcal{L}_{\textrm{Sym}}&=&i\psi^T (\partial_t-\bar{v}_n \partial_x)\psi +\psi^T(\xi_{ab}L^{ab})\psi; \nonumber \\
\mathcal{L}_{\textrm{SB}}&=&-i\psi^T (\delta v \partial_x) \psi -v_{\rho n_1} (\partial_x \phi_{\rho}) \psi^T M\psi,
\end{eqnarray}
where $\psi^T\equiv(\psi_1,\psi_2,\psi_3,\psi_4)$ and $a,b=1,2,3,4$. The new random variables are defined as $\xi_{13}=-2\pi (\textrm{Im}\xi_{1}+\textrm{Im}\xi_{2})$, $\xi_{14}=-2\pi (\textrm{Re}\xi_{1}-\textrm{Re}\xi_{2})$, $\xi_{23}=2\pi (\textrm{Re}\xi_{1}+\textrm{Re}\xi_{2})$, $\xi_{24}=-2\pi (\textrm{Im}\xi_{1}-\textrm{Im}\xi_{2})$, and all other $\xi_{ab}=0$. The Hermitian matrices $L^{ab}$ are the generators of the $SO(4)$ group in the fundamental representation,
\begin{equation}
\label{dimaA10}
(L^{ab})_{kl}=i(\delta_{al}\delta_{bk}-\delta_{ak}\delta_{bl})
\end{equation}
with $k,l=1,2,3,4$. The other two matrices in the Lagrangians are $\delta v=\textrm{Diag}(\delta v_{n_1}, \delta v_{n_1},\delta v_{n_2},\delta v_{n_2})$ and $M=-L^{12}$. 

As the last step, we perform a local rotation of the Majorana fermions $\tilde{\psi}(x)=R(x)\psi(x)$, where
\begin{equation}
\label{dimaA11}
R(x)=\mathcal{P} \exp \Big( \frac{i}{\bar{v}_n}\int_{-\infty}^x dx' \xi_{ab}(x') L^{ab} \Big)
\end{equation}
and $\mathcal{P}$ is the path-ordering operator. Clearly, $R^TR=\mathbb{1}_{4\times4}$ and $\det(R)=1$. After the rotation, the Lagrangians become
\begin{eqnarray}
\label{dimaA12}
\mathcal{L}_{\textrm{Sym}}&=&i\tilde{\psi}^T (\partial_t-\bar{v}_n \partial_x)\tilde{\psi}; \nonumber \\
\mathcal{L}_{\textrm{SB}}&=&-i\tilde{\psi}^T (\widetilde{\delta v} \partial_x) \tilde{\psi} -v_{\rho n_1}(\partial_x \phi_{\rho}) \tilde{\psi}^T \widetilde{M} \tilde{\psi},
\end{eqnarray}
where $\widetilde{\delta v}=R(x)\delta vR^{-1}(x)$ and $\widetilde{M}=R(x)MR^{-1}(x)$ are spatially random matrices. 
On the second line of the above equation we omit a term with the derivative $\partial_x R^{-1}$. Its structure is similar to the second term on the first line of Eq. (\ref{dimaA9}) and it can be removed with  a variable change,
similar to Eq. (\ref{dimaA11}).
The $SO(4)$ symmetry in $\mathcal{L}_{\textrm{Sym}}$ is now clearly manifest. Let us study the symmetry-breaking terms in $\mathcal{L}_{\textrm{SB}}$. Naively, both operators in $\mathcal{L}_{\textrm{SB}}$ have scaling dimension 2 and they are marginal. However, the random coefficients  $\widetilde{\delta v}$ and $\widetilde{M}$ make them irrelevant under a perturbative RG analysis.\cite{39} Hence,  both terms in $\mathcal{L}_{\textrm{SB}}$ scale to zero at low temperature.

In general, our initial simplifying assumptions for the potential matrix may not hold and we will have two more terms in the Lagrangian,
\begin{align}
\label{dimaA13}
\mathcal{L}'_{\textrm{Sym}}=&-\frac{1}{4\pi}v_{n_1n_2}\partial_{x}\phi_{n_1}\partial_{x}\phi_{n_2} =\frac{\pi }{6}v_{n_1n_2}\varepsilon^{ijkl}\tilde{\psi}_i\tilde{\psi}_j\tilde{\psi}_k\tilde{\psi}_l; \nonumber \\
\mathcal{L}'_{\textrm{SB}}= &-\frac{1}{4\pi}v_{\rho n_2}\partial_{x}\phi_{\rho}\partial_{x}\phi_{n_{2}} =-v_{\rho n_2}(\partial_x \phi_{\rho}) \tilde{\psi}^T N \tilde{\psi},
\end{align}
where $\varepsilon^{ijkl}$ is the Levi-Civita tensor with $i,j,k,l=1,2,3,4$, and $N$ is a spatially random matrix. $\mathcal{L}'_{\textrm{Sym}}$ is a marginal term that respects the $SO(4)$ symmetry. It should be retained in the edge theory of the anti-331 state. On the other hand, the operator in $\mathcal{L}'_{\textrm{SB}}$ is irrelevant.

Any $SO(4)$-invariant combination of the Majorana fermions can be constructed from the Majorana operators and their derivatives, the Kronecker tensor $\delta_{ij}$ and the Levi-Civita tensor $\varepsilon^{ijkl}$. 
Keeping all possible relevant terms we find
 the full Lagrangian that describes the low energy edge physics of the anti-331 state:
\begin{align}
\label{dimaA14}
\mathcal{L}=&-\frac{2}{4\pi} [\partial_{t}\phi_{\rho}\partial_{x}\phi_{\rho}+v_{\rho} (\partial_{x}\phi_{\rho})^{2}]+i\tilde{\psi}^T (\partial_t-\bar{v}_n \partial_x)\tilde{\psi}\nonumber \\
& +\frac{\pi }{6}v_{n_1n_2}\varepsilon^{ijkl}\tilde{\psi}_i\tilde{\psi}_j\tilde{\psi}_k\tilde{\psi}_l.
\end{align}
The edge theory consists of one right-moving charged density mode and four left-moving Majorana fermions obeying an explicit $SO(4)$ symmetry. The density mode and the Majorana fermions decouple.

The thermal conductance of a FQH state is a universal quantity that is determined by the bulk topological order. It depends only on the numbers of right-moving and left-moving edge modes and is robust with respect to interactions (for example, the off-diagonal elements in the potential matrix) and disorder.\cite{40} Using the Lagrangian in Eq.~(\ref{dimaA1}) or Eq.~(\ref{dimaA14}), we obtain the thermal conductance in the anti-331 state as $\kappa_{\overline{331}}=2+1-1-1=2+ 1-4\times \frac{1}{2}=1$, in units of $\pi^2 k_B^2T/3h$ (the first factor of 2 comes from the two integer modes). This is different from the thermal conductance $\kappa_{331}=2+ 1+1=4$ in the 331 state, indicating different topological orders of the 331 and  anti-331 state.

A generic quasiparticle operator on the edge of the anti-331 state is $\Psi_{qp}=e^{i (l_0 \phi_0+l_1 \phi_1+l_2 \phi_2)}$ with the electric charge $Q_{qp}=l_0-\frac{1}{2}l_1-\frac{1}{4}l_2$ in units of $e$, where $l_0$, $l_1$, $l_2$ are independent integers. Its scaling dimension $g_{qp}$ can be obtained by computing the correlation function, $\langle \Psi_{qp}^{\dagger}(t)\Psi_{qp}(0)\rangle  \sim t^{-2g_{qp}}$ at zero temperature, using the $SO(4)$ invariant Lagrangian from Eq.~(\ref{dimaA14}). Equivalently yet more easily, we can use the boson Lagrangians from Eqs.~(\ref{dimaA7},\ref{dimaA8},\ref{dimaA13}) by setting $\delta v_1=\delta v_2 =v_{\rho n_1}=v_{\rho n_2}=0$ and $\xi_{1}=\xi_2=0$. This is because the rotation $R(x)$ is local and does not change the equal-space correlation functions of quasiparticle operators. The interaction $v_{n_1n_2}$ between copropagating neutral bosons cannot modify the scaling behavior of a quasiparticle \cite{wen}. Using $\phi_{0}=2\phi_{\rho}-\phi_{n_1}$, $\phi_{1}=-\phi_{\rho}+\phi_{n_1}$ and $\phi_{2}=-\frac{1}{2}(\phi_{\rho}-\phi_{n_1}-\phi_{n_2})$, we obtain the universal scaling dimension of $\Psi_{qp}$ as
\begin{equation}
\label{dimaA15}
g_{qp}=\frac{(4l_0-2l_1-l_2)^2}{16}+\frac{(2l_0-2l_1-l_2)^2}{8}+\frac{l_2^2}{8}.
\end{equation}
The most relevant quasiparticles are those with the minimal scaling dimension (\ref{dimaA15}). Here, it is $e^{i ( \phi_0+ \phi_1)}$ with the electric charge $e/2$ and scaling dimension $1/4$. The next-most relevant quasiparticles are $e^{i ( \phi_0+ 2\phi_1-\phi_2)}$, $e^{i ( \phi_0+ \phi_1+ \phi_2)}$, $e^{i (\phi_2-\phi_1)}$ and $e^{-i \phi_2}$ with the electric charge $e/4$ and scaling dimension $5/16$. The above quasiparticle operators can all be represented in the form $e^{i\phi_\rho/2   }e^{\pm i\phi_{n_1}/2}e^{\pm i \phi_{n_2}/2}$, where the last two exponential factors correspond to the operators $\sigma_\alpha$ in Section II.F.

\section{Edge of the anti-$SU(2)_{2}$ state}
In this appendix we discuss the edge physics of the anti-$SU(2)_{2}$ state, which is the disorder-dominated particle-hole conjugate of the $SU(2)_{2}$ state. 
As in the previous appendix, we focus on the FQH edge separating $\nu=2$ and $\nu=5/2$ regions.

Our approach is the same as in Appendix A.
As a starting point, we write down the edge Lagrangian of a clean particle-hole conjugate of the $SU(2)_{2}$ phase
\begin{align}
\label{dimaB1-new}
\mathcal{L}_0= & -\frac{1}{4\pi}  [\partial_{t}\phi_{0}\partial_{x}\phi_{0}-2\partial_{t}\phi_{\rho}\partial_{x}\phi_{\rho}-\partial_{t}\phi_{n}\partial_{x}\phi_{n} \nonumber \\ 
&+v_{0}(\partial_{x}\phi_{0})^{2}+2v_{\rho} (\partial_{x}\phi_{\rho})^{2} +v_{n} (\partial_{x}\phi_{n})^{2}]\nonumber \\ 
&-\frac{2}{4\pi}v_{0\rho}\partial_{x}\phi_{0}\partial_{x}\phi_{\rho}+i \psi (\partial_{t}-v_{\psi}\partial_{x})\psi ,
\end{align}
where $\phi_{0}$ is the right-moving $\nu=1$ edge mode and $\phi_{\rho}$, $\phi_{n}$, $\psi$ are left-moving $SU(2)_{2}$ FQH edge modes. We first assume that only charged modes interact via Coulomb potential $v_{0\rho}$.
At the end of the appendix we will see that our conclusions do not depend on that assumption.
The charge vector of the boson fields ($\phi_0,\phi_{\rho},\phi_n$) is $\mathbf{t}^T=(1,1,0)$. The Majorana fermion $\psi$ is neutral.

In the presence of disorder, electrons may tunnel between the integer edge and the FQH edge. The electron operator on the $\nu=1$ edge is $e^{i\phi_0}$. The most relevant electron operators on the FQH edge are $\psi e^{-i2\phi_{\rho}}$, $\textrm{Re}\{e^{-i\phi_n}\} e^{-i2\phi_{\rho}}$ and $\textrm{Im} \{e^{-i\phi_n}\} e^{-i2\phi_{\rho}}$. The Lagrangian density that describes the electron tunneling due to impurities on the edge is
\begin{align}
\label{dimaB_add}
\mathcal{L}_{\textrm{tun}}=&\xi_{1}(x) \psi e^{i(\phi_0+2\phi_{\rho})}+\xi_{2}(x) \textrm{Re}\{e^{-i\phi_n}\} e^{i(\phi_0+2\phi_{\rho})} \nonumber \\
&+ \xi_{3}(x) \textrm{Im}\{e^{-i\phi_n}\} e^{i(\phi_0+2\phi_{\rho})} + \textrm {H.c.},
\end{align}
where $\xi_{i}(x)$ are complex Gaussian variables characterizing the strength of disorder, $\langle \xi_{i}(x)\xi_{j}^*(x')\rangle =W_i \delta(x-x')\delta_{ij}$, where $i,j=1,2,3$. To simplify notation we suppress Klein factors as in Appendix A.

From $\mathcal{L}_0$ in Eq.~(\ref{dimaB1-new}), we find that the tunneling operators in Eq. (\ref{dimaB_add}) have the same scaling dimension $\Delta=1/2+(3-2\sqrt{2}c)/(2 \sqrt{1-c^2})$, where $c=\sqrt{2} v_{0\rho}/(v_{0}+v_{\rho})$. The scaling dimension $\Delta$ determines the edge physics at low temperatures. The leading order RG equations are $dW_i/dl=(3-2\Delta)W_i$.\cite{39} When $\Delta>3/2$, the electron tunneling is irrelevant and the low temperature edge physics is described by $\mathcal{L}_0$. When $\Delta<3/2$, we arrive at the disorder-dominated phase with equilibrated edges at low temperature.

The physics of the non-equilibrated case $\Delta<3/2$ is nonuniversal. Just like in Appendix C we find that the quasiparticle tunneling exponent $g$ can assume a broad range of values. Moreover, the quantum Hall conductance at low temperatures and voltages does not assume the correct value $5e^2/2h$. It becomes $7e^2/2h$ in the quantum Hall bar geometry. This happens for the same reasons as in Appendices A and C: One has to add the conductance $3e^2/h$ of the integer quantum Hall subsystem and the conductance $e^2/2h$ of the FQH subsystem. Below we focus on the disorder dominated equilibrated phase.

Let us separate the charged and neutral degrees of freedom on the edge by defining $\phi_{\tilde{\rho}}=\phi_0+\phi_{\rho}$, $\phi_{n_1}=\phi_0+2\phi_{\rho}$ and $\phi_{n_2}=\phi_n$. The full Lagrangian $\mathcal{L}_0+\mathcal{L}_{\textrm{tun}}=\mathcal{L}_{\tilde{\rho}}+\mathcal{L}_{\textrm{Sym}}+\mathcal{L}_{\textrm{SB}}$, where
\begin{align}
\label{dimaB2}
\mathcal{L}_{\tilde{\rho}}=&-\frac{2}{4\pi} [\partial_{t}\phi_{\tilde{\rho}} \partial_{x}\phi_{\tilde{\rho}}+v_{\tilde{\rho}} (\partial_{x}\phi_{\tilde{\rho}})^{2}]; \nonumber \\
\mathcal{L}_{\textrm{Sym}}=& -\frac{1}{4\pi} \sum_{i=1,2} [-\partial_{t}\phi_{n_{i}}\partial_{x}\phi_{n_{i}}+\bar{v}_{n} (\partial_{x}\phi_{n_{i}})^{2}] +[\xi_{1} \psi e^{i\phi_{n_{1}}} \nonumber \\
&+\xi_{2}\textrm{Re}\{e^{-i\phi_{n_2}}\} e^{i\phi_{n_{1}}} + \xi_{3} \textrm{Im}\{e^{-i\phi_{n_2}}\} e^{i\phi_{n_{1}}} + \textrm {H.c.}] \nonumber \\
&+ i \psi (\partial_{t}- \bar{v}_n\partial_{x})\psi; \nonumber \\
\mathcal{L}_{\textrm{SB}}=& -\frac{1}{4\pi} \sum_{i=1,2}\delta v_{n_i} (\partial_{x}\phi_{n_{i}})^{2}-\frac{2}{4\pi}v_{\tilde{\rho} n_1}\partial_{x}\phi_{\tilde{\rho}}\partial_{x}\phi_{n_{1}}\nonumber \\
&-i \psi (\delta v_{\psi} \partial_{x})\psi.
\end{align} 
The velocities of $\phi_{\tilde{\rho}}$, $\phi_{n_1}$, $\phi_{n_2}$ are $v_{\tilde{\rho}}=2v_0+v_{\rho}-2v_{0\rho}$, $v_{n_1}=v_0+2v_{\rho}-2v_{0\rho}$ and $v_{n_2}=v_{n}$, respectively. The interaction parameter $v_{\tilde{\rho} n_1}=-2v_0-2v_{\rho}+3v_{0\rho}$. Here we have split the Lagrangian into $\mathcal{L}_{\textrm{Sym}}$, which respects a hidden symmetry, and the symmetry-breaking part $\mathcal{L}_{\textrm{SB}}$. The velocity anisotropies $\delta v_{\psi}= v_{\psi}- \bar{v}_{n}$ and $\delta v_{n_i}= v_{n_i}- \bar{v}_{n}$ ($i=1,2$) are defined with respect to the average velocity $\bar{v}_{n}=\frac{1}{5}(v_{\psi}+2v_{n_1}+2v_{n_2})$. 

We now fermionize the neutral bosons into Majorana fermions according to $\psi_2=\frac{1}{\sqrt{2 \pi}}\textrm{Re}\{ e^{-i\phi_{n_1}}\}$, $\psi_3=\frac{1}{\sqrt{2 \pi}}\textrm{Im}\{e^{-i\phi_{n_1}}\}$, $\psi_4= \frac{1}{\sqrt{2 \pi}}\textrm{Re}\{e^{-i \phi_{n_2}}\}$, $\psi_5=\frac{1}{\sqrt{2 \pi}}\textrm{Im}\{e^{-i\phi_{n_2}}\}$ and define $\psi_1=\psi$. In terms of the Majorana fermions,
\begin{eqnarray}
\mathcal{L}_{\textrm{Sym}}&=&i\psi^T (\partial_t-\bar{v}_n \partial_x)\psi +\psi^T(\xi_{ab}K^{ab})\psi \nonumber \\
\mathcal{L}_{\textrm{SB}}&=&-i\psi^T (\delta v \partial_x) \psi -v_{\tilde{\rho} n_1} (\partial_x \phi_{\tilde{\rho}}) \psi^T M\psi,
\end{eqnarray}
where $\psi^T\equiv(\psi_1,\psi_2,\psi_3,\psi_4,\psi_5)$ and $a,b=1,2,3,4,5$. Random variables $\xi_{ab}$ are defined as $\xi_{12}=-\sqrt{2 \pi} \textrm{Im}\xi_{1}$, $\xi_{13}=\sqrt{2 \pi} \textrm{Re}\xi_{1}$, $\xi_{24}=2 \pi \textrm{Im}\xi_{2}$, $\xi_{25}=2 \pi \textrm{Im}\xi_{3}$, $\xi_{34}=-2 \pi \textrm{Re}\xi_{2}$, $\xi_{35}=- 2 \pi \textrm{Re}\xi_{3}$ and all other $\xi_{ab}=0$. Hermitian matrices $K^{ab}$ are the generators of the $SO(5)$ group in the fundamental representation, $(K^{ab})_{kl}=i(\delta_{al}\delta_{bk}-\delta_{ak}\delta_{bl})$ with $k,l=1,2,3,4,5$. The other matrices in the Lagrangians are $\delta v=\textrm{Diag}(\delta v_{\psi},\delta v_{n_1}, \delta v_{n_1},\delta v_{n_2},\delta v_{n_2})$ and $M=-K^{23}$. 

Finally, we make a local rotation of the Majorana fermions $\tilde{\psi}(x)=R(x)\psi(x)$, where
\begin{equation}
\label{dimaR-new}
R(x)=\mathcal{P} \exp \Big( \frac{i}{\bar{v}_n}\int_{-\infty}^x dx' \xi_{ab}(x') K^{ab} \Big)
\end{equation}
and $\mathcal{P}$ is the path-ordering operator. The Lagrangians after the rotation are\begin{eqnarray}
\label{dimaB1}
\mathcal{L}_{\textrm{Sym}}&=&i\tilde{\psi}^T (\partial_t-\bar{v}_n \partial_x)\tilde{\psi}  \nonumber \\ 
\mathcal{L}_{\textrm{SB}}&=&-i\tilde{\psi}^T (\widetilde{\delta v} \partial_x) \tilde{\psi} -v_{\tilde{\rho} n_1}(\partial_x \phi_{\tilde{\rho}}) \tilde{\psi}^T \widetilde{M} \tilde{\psi},
\end{eqnarray}
where $\widetilde{\delta v}=R(x)\delta vR^{-1}(x)$ and $\widetilde{M}=R(x)MR^{-1}(x)$ are spatially random matrices.
As in Appendix A, we omit a term with $\partial_x R^{-1}$. It can be removed with another transformation of the type (\ref{dimaR-new}).
 Now, the $SO(5)$ symmetry in $\mathcal{L}_{\textrm{Sym}}$ is clear. The symmetry-breaking terms in $\mathcal{L}_{\textrm{SB}}$ are irrelevant due to their random coefficients, in exactly the same way as in the anti-331 state, Appendix A. 

In the most general case, the neutral boson $\phi_n$ in Eq.~(\ref{dimaB1-new}) may interact with charged bosons $\phi_0$ and $\phi_{\rho}$. This gives rise to two more symmetry-breaking terms in the Lagrangian. However, both terms are irrelevant and disappear at low temperatures. The edge physics of the anti-$SU(2)_{2}$ state is described by $\mathcal{L}_{\tilde{\rho}}+\mathcal{L}_{\textrm{Sym}}$ that contains all relevant operators allowed by the symmetry.
In contrast to the anti-331 case with its $SO(4)$ symmetry group, we do not need to keep any four-fermion operators in the action.

A generic quasiparticle operator on the edge of the anti-$SU(2)_{2}$ state is $\Psi_{qp}=\zeta e^{i (l_0 \phi_0+l_{\rho} \phi_{\rho}+l_n \phi_n)}$ with the electric charge $Q_{qp}=l_0-\frac{1}{2}l_{\rho}$ in units of $e$, where $\zeta$ is one of the three fields $\mathbf{1}$, $\sigma$, and $\psi$ with the scaling dimensions $g_{\mathbf{1}}=0$, $g_{\sigma}=\frac{1}{16}$ and $g_{\psi}=\frac{1}{2}$. $l_0$ is an arbitrary integer. $l_{\rho}$ and $l_n$ are integers or half integers. We are interested in $\pm e/4$ charged quasiparticles. The quasiparticle charge gives the constraint $l_0=\frac{1}{2} l_{\rho} \pm \frac{1}{4}$, which is satisfied only if $l_{\rho}$ is a half integer given that $l_0$ is an integer. In addition, we must require that the quasiparticle is local with respect to electrons, which means there is no branch cut in the correlation function between the quasiparticle operator and any of the electron operators in the theory \cite{wen}. Hence, $l_n$ must also be a half integer and $\zeta=\sigma$. The most general $\pm e/4$ charged quasiparticle is then $\Psi_{\pm e/4}=\sigma e^{i (l_0 \phi_0+l_{\rho} \phi_{\rho}+l_n \phi_n)}$ with $l_{\rho}$ and $l_n$ being  half integers and $l_0=\frac{1}{2} l_{\rho} \pm \frac{1}{4}$ an integer. The scaling dimension $g_{\pm e/4}$ can be computed using the boson Lagrangians in Eq.~(\ref{dimaB2}) by setting $\delta v_{\psi}=\delta v_1=\delta v_2 =v_{\tilde{\rho} n_1}=0$ and $\xi_{1}=\xi_2=\xi_3=0$. This gives
\begin{equation}
g_{\pm e/4}=\frac{1}{2} (l_0 \mp \frac{1}{2})^2+\frac{1}{2} l_n^2 +\frac{1}{8}.
\end{equation}
The most relevant $e/4$ and $-e/4$ charged quasiparticles are $\sigma e^{i ( \phi_0+ \frac{3}{2} \phi_{\rho} \pm \frac{1}{2} \phi_n)}$, $\sigma e^{-i (\frac{1}{2} \phi_{\rho} \pm \frac{1}{2} \phi_n)}$ and $\sigma e^{-i ( \phi_0+ \frac{3}{2} \phi_{\rho} \pm \frac{1}{2} \phi_n)}$, $\sigma e^{i (\frac{1}{2} \phi_{\rho} \pm \frac{1}{2} \phi_n)}$, all with the scaling dimension $3/8$. All these operators are products of a factor $e^{\pm i\phi_{\tilde\rho}/2}$ and one of the twist operators $\sigma_\alpha$, Section II.H.

\section{Edge of the anti-$K=8$ state}
The anti-$K=8$ state is the particle-hole conjugate of the $K=8$ state, with the FQH edge Lagrangian density 
\begin{align}
\mathcal{L}_0= & -\frac{1}{4\pi}  [\partial_{t}\phi_{0}\partial_{x}\phi_{0}-8\partial_{t}\phi_{1}\partial_{x}\phi_{1}+v_{0}(\partial_{x}\phi_{0})^{2} \nonumber \\
&+8v_{1} (\partial_{x}\phi_{1})^{2}] -\frac{1}{\pi}v_{01}\partial_{x}\phi_{0}\partial_{x}\phi_{1},
\end{align}
where $\phi_0$ is the right-moving integer edge mode and $\phi_1$ is the left-moving fractional edge mode from the $K=8$ state. 
As in Appendices A and B we focus on the FQH edge between the $\nu=2$ and $\nu=5/2$ regions.
The charge vector of ($\phi_0,\phi_1$) is $\mathbf{t}^T=(1,2)$, which reflects that the $K=8$ state is a Laughlin state of electron pairs. The Lagrangian that describes the tunneling of electron pairs between the integer and fractional edges due to impurities is
\begin{equation}
\mathcal{L}_{\textrm{tun}}= \xi(x) e^{i(2\phi_{0}+8\phi_{1})}+  \textrm {H.c.},
\end{equation}
where $\xi(x)$ is a complex Gaussian random variable that describes local disorder, $\langle \xi(x)\xi^*(x')\rangle =W \delta(x-x')$. The scaling dimension of the tunneling operators is $\Delta=(6-4\sqrt{2}c)/\sqrt{1-c^2}$, where $c=\sqrt{2} v_{01}/(v_{0}+v_{1})$. Positive-definiteness of the Hamiltonian requires $v_0v_1>v_{01}^2/2$, and hence $|c|<1$. The leading order RG equation for the disorder strength $W$ is $dW/dl=(3-2\Delta)W$.\cite{39} By inspection, the scaling dimension $\Delta$ is always greater than $3/2$. Hence, electron-pair tunneling is always irrelevant and a disorder-dominated phase does not exist. The edge physics of the anti-$K=8$ state is described by $\mathcal{L}_0$.

A generic quasiparticle $\Psi_{qp}= e^{i (l_0 \phi_0+l_{1} \phi_{1})}$ ($l_0$, $l_1$ are independent integers) on the edge of the anti-$K=8$ state has the electric charge $Q_{qp}=l_0-\frac{1}{4}l_{1}$ in units of $e$. Its scaling dimension $g_{qp}$ is nonuniversal and depends on the parameters in the Hamiltonian. With $\mathcal{L}_0$, we find
\begin{equation}
g_{qp}=\frac{1}{2\sqrt{1-c^2}}(l_0^2+\frac{l_1^2}{8}-\frac{l_0l_1}{\sqrt{2}}c).
\end{equation}
Possible values of $g_{qp}$ for  charge-$e/4$ excitations range between $1/16$ and $+\infty$. The low temperature quantum Hall bar conductance is quantized at $7e^2/2h$ just like in 
other nonequilibrated states with the same filling factor, Ref. \onlinecite{wang10a}. 

The above discussion ignores the two integer edge modes $\phi^0_1$, $\phi^0_2$ always present in the second Landau level states. If we allow tunneling between those modes and the FQH modes and include operators that transmit three or more electrons then it is possible to find relevant tunneling operators at certain choices of the interaction constants in the Hamiltonian. One example would be the operator $O_3=\exp(i[8\phi_1+3\phi_0-\phi^0_1])$, where $\phi^0_1$ describes the integer mode with the same spin polarization as the FQH edge. We expect the amplitude of such many-body operators to be small and neglect them even if the interaction constants are such that they are relevant.

\section{Positive-definiteness of the interaction matrix $U_{ij}$}

\begin{figure}
\centering
\includegraphics[width=2.7in]{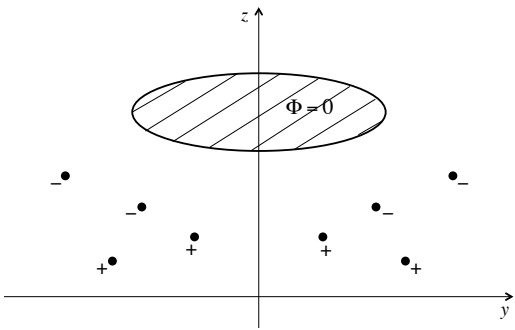}
\caption{The charge distribution is mirror symmetric with respect to the $y=0$ plane. The mirror-symmetric gate with the potential $\Phi=0$ is shaded.}
\end{figure}

In this appendix we prove that the interaction matrix $U_{ij}$ in Eq. (\ref{dima16}) is positive definite. We consider a system with mirror symmetry, Fig. 6. In Fig. 6 the $x$-axis is perpendicular to the plane of the figure and the mirror plane is defined by the equation $y=0$. The only assumption about the dielectric constant outside the mirror symmetric gates is the mirror symmetry of its coordinate dependence.
  Our effective models for edge states are one-dimensional and contain only the coordinate $x$ but the physical charge distribution is always three-dimensional. The energy of the electrostatic interaction between the edge segments on the left and right of the gate, Fig. 6, depends on the whole three-dimensional distribution of the charges. 

The derivatives of the Bose fields $\partial_x\phi_i$ in Eq. (\ref{dima16}) are determined by the local charge distribution. It will be convenient for us to consider a situation in which the charge density $\rho({\bf r})=\rho(x,y,z)$ does not depend on $x$ and exhibits the mirror symmetry $\rho(x,y,z)=\rho(x,-y,z)$.
Then all $\partial_x\phi_i$ remain constant on the left and right of the gate with $\partial_x\phi_i({\rm left ~of~the ~gate})=-\partial_x\phi_i({\rm right~of ~the~ gate})$. The interaction energy (\ref{dima16}) becomes $(+1/4\pi)\int_{-\infty}^0 dx
\sum_{ij}U_{ij}\partial_x\phi_i(x<0)\partial_x\phi_j(x<0)$. The same energy can be found from electrostatics. It is just the interaction energy of the charge distribution on the left of the gate, $\rho(x,y,z)$, $y<0$, with the mirror symmetric charge distribution $\rho(x,y,z)$, $y>0$. 
It is now clear that to prove the positive definiteness of $U$
it is sufficient to prove that the electrostatic interaction energy of a set of charges on the left of the mirror plane with the mirror symmetric set of charges on the right of the plane is always positive. Below we compute such electrostatic energy in the presence of mirror symmetric metallic gates.

We set the electrostatic potential of the gates to zero.
The electrostatic potential $\Phi({\bf r})$ outside the gates  can be represented as the sum of two contributions. $\Phi_l({\bf r})$ is the electrostatic potential, created by the charges on the left of the mirror plane (i.e., the charges from the points with negative $y$). By definition $\Phi_l$ includes the effect of the screening charges on the gate surface, i.e., satisfies the boundary condition $\Phi_l({\bf r}{\rm~ in~ the~ gate})=0$.
$\Phi_r({\bf r})$ is created by the charges on the right of the mirror plane in the presence of a screening gate. The mirror symmetry implies that $\Phi_l(x,y,z)=\Phi_r(x,-y,z)$. The total electrostatic energy is thus

\begin{equation}
\label{dimaD1}
E=\int dxdz\int_{-\infty}^0 dy \rho(x,y,z)[\Phi_l(x,y,z)+\Phi_r(x,y,z)].
\end{equation}
We wish to prove that the interaction contribution 

\begin{equation}
\label{dimaD2}
E_{lr}=\int dxdz\int_{-\infty}^0 dy \rho \Phi_r
\end{equation}
is positive. 

Let us investigate the effect of two changes in the charge distribution on the energy.

1) We remove all charges on the right of the mirror plane, i.e., set $\rho(x,y>0,z)=0$. The electrostatic energy becomes

\begin{equation}
\label{dimaD3}
E_1=\int dxdz\int_{-\infty}^0\rho(x,y,z)\Phi_l(x,y,z)/2.
\end{equation}

2) Alternatively, let us change the sign of all charges on the right of the mirror plane: $\rho(x,y,z)\rightarrow-\rho(x,y,z)$, $y>0$. The potential becomes $\Phi_l({\bf r})-\Phi_r({\bf r})$.
We have created a situation with $\Phi(x,y=0,z)=0$. Hence, the total electrostatic potential at $y<0$ would not change after all charges with $y>0$ are removed and the region $y>0$ is filled with metal. In the latter situation, the total electrostatic energy becomes

\begin{equation}
\label{dimaD4}
E_2=\int dxdz\int_{-\infty}^0 dy \rho(x,y,z)[\Phi_l(x,y,z)-\Phi_r(x,y,z)]/2,
\end{equation}
where $\Phi_l$ and $\Phi_r$ are the same as in Eq. (\ref{dimaD1}), i.e., $[\Phi_l(x,y,z)+\Phi_r(x,y,z)]$ is the electrostatic potential of a mirror symmetric charge distribution.

According to a theorem of electrostatics \cite{LL}, the energy always goes down if a piece of metal is introduced into a system at fixed positions of free charges. Hence, $E_2<E_1$. Subtracting $E_2$ from $E_1$ we find that $E_{lr}>0$ and hence $U_{ij}$ is positive definite.


\end{document}